# Asteroid Ryugu Before the Hayabusa2 Encounter

Koji Wada[1], Matthias Grott[2], Patrick Michel[3], Kevin J. Walsh[4], Antonella M. Barucci[5], Jens Biele[6], Jürgen Blum[7], Carolyn M. Ernst[8], Jan T. Grundmann[9], Bastian Gundlach[7], Axel Hagermann[10], Maximilian Hamm[2], Martin Jutzi[11], Myung-Jin Kim[12], Ekkehard Kührt[2], Lucille Le Corre[13], Guy Libourel[3], Roy Lichtenheldt[14], Alessandro Maturilli[2], Scott R. Messenger[15], Tatsuhiro Michikami[16], Hideaki Miyamoto[17], Stefano Mottola[2], Akiko M. Nakamura[18], Thomas Müller[19], Larry R. Nittler[20], Kazunori Ogawa[18], Tatsuaki Okada[21], Ernesto Palomba[22], Naoya Sakatani[21], Stefan Schröder[2], Hiroki Senshu[1], Driss Takir[23], Michael E. Zolensky[15], and International Regolith Science Group (IRSG) in Hayabusa2 project

***Corresponding Author:***

Koji Wada

Planetary Exploration Research Center (PERC),

Chiba Institute of Technology (Chitech),

Tsudanuma 2-17-1, Narashino, Chiba 275-0016, Japan

Email: wada@perc.it-chiba.ac.jp

tel : +81-47-478-4744, fax : +81-47-478-0372

***Affiliations***:

[1] Planetary Exploration Research Center (PERC), Chiba Institute of Technology, Chiba, Japan

[2] DLR Institute for Planetary Research, Rutherfordstr. 2, 12489 Berlin, Germany

[3] Université Côte d'Azur, Observatoire de la Côte d'Azur, CNRS, Lagrange Laboratory

[4] Southwest Research Institute, Boulder, USA

[5] Laboratoire d'Etudes Spatiales et d'Instrumentation en Astrophysique (LESIA), Observatoire de Paris, PSL Research University, CNRS, Université Paris Diderot, Sorbonne Paris Cité, UPMC Université Paris 06, Sorbonne Universités, 5 Place J. Janssen, Meudon Principal Cedex 92195, France

[6] DLR RB-MSC German Aerospace Center, 51147 Cologne, Germany

[7] Institut für Geophysik und extraterrestrische Physik, Technische Universität Braunschweig, Mendelssohnstr. 3, D-38106 Braunschweig, Germany

[8] Johns Hopkins University Applied Physics Laboratory, Laurel, MD, USA

[9] DLR German Aerospace Center, Institute of Space Systems, System Engineering and Project Office, Robert-Hooke-Strasse 7, D-28359 Bremen, Germany

[10] Department of Biological and Environmental Sciences, University of Stirling, FK9 4LA, Scotland

[11] Physics Institute, University of Bern, Switzerland

[12] Center for Space Situational Awareness, Korea Astronomy and Space Science Institute, 776, Daedeokdae-ro, Yuseong-gu, Daejeon, 305-348, Republic of Korea

[13] Planetary Science Institute, 1700 East Fort Lowell, Suite 106, Tucson, AZ 85719, USA

[14] DLR Institute of System Dynamics and Control, Oberpfaffenhoffen, Germany

[15] Robert M Walker for Space Sciences, Astromaterials Research and Exploration Science, NASA Johnson Space Center, Houston, TX 77058, USA

[16] Faculty of Engineering, Kindai University, Hiroshima Campus, 1 Takaya Umenobe, Higashi-Hiroshima, Hiroshima 739-2116, Japan

[17] Deptartment of Systems Innovation, University of Tokyo, Tokyo, Japan

[18] Department of Planetology, Graduate School of Science, Kobe University, 1-1 Rokkodai, Nada-ku, Kobe 657-8501, Japan

[19] Max-Planck Institute for Extraterrestrial Physics, Garching, Germany

[20] Department of Terrestrial Magnetism, Carnegie Institution of Washington, 5241 Broad Branch Rd NW, 20015, USA

[21] Institute of Space and Astronautical Science, Japan Aerospace Exploration Agency, Sagamihara, Japan

[22] INAF, Istituto di Astrofisica e Planetologia Spaziali, via Fosso del Cavaliere, 00133 Rome, Italy

[23] SETI Institute, 189 Bernardo Ave., Mountain View, CA 94043, USA

### *Abstract*

Asteroid (162173) Ryugu is the target object of Hayabusa2, an asteroid exploration and sample return mission led by Japan Aerospace Exploration Agency (JAXA). Ground-based observations indicate that Ryugu is a C-type near-Earth asteroid with a diameter of less than 1 km, but the knowledge of its detailed properties is still very limited. This paper summarizes our best understanding of the physical and dynamical properties of Ryugu based on remote sensing and theoretical modeling. This information is used to construct a design reference model of the asteroid that is used for formulation of mission operations plans in advance of asteroid arrival. Particular attention is given to the surface properties of Ryugu that are relevant to sample acquisition. This reference model helps readers to appropriately interpret the data that will be directly obtained by Hayabusa2 and promotes scientific studies not only for Ryugu itself and other small bodies but also for the Solar System evolution that small bodies shed light on.

# Table of Contents

# 1. Introduction

Asteroids and comets are relicts of the early stages of solar system evolution. They contain materials and structures which are relatively primitive compared to those composing planets and thus provide a window into the earliest stages of planet formation. Investigation of various properties of small bodies thus provides crucial knowledge needed to understand the origin and evolution of the solar system. The asteroid (162173) Ryugu is a C-type near-Earth asteroid (NEA) with a diameter of less than 1 km. It is the target of Hayabusa2, the second asteroid sample return mission led by Japan Aerospace Exploration Agency (JAXA) (Tsuda et al. 2013). Ryugu will offer a great opportunity to understand the present status and the evolutionary history of volatile-rich primitive materials in the solar system (Watanabe et al. 2017). This expectation is supported by the fact that Hayabusa, the previous and first asteroid exploration and sample return mission by JAXA, returned a lot of surprising and fruitful data from remote-sensing observations and the returned particles of the S-type NEA (25143) Itokawa (e.g., Fujiwara et al. 2006; Miyamoto et al. 2007; Nagao et al. 2011; Nakamura et al. 2011; Yoshikawa et al. 2015). Although small bodies are full of treasures, only a limited number have been visited by spacecraft so far, such as asteroid (433) Eros (e.g., Cheng 2002) and comet 67P/Churyumov-Gerasimenko (e.g., Sierks et al. 2015). C-type (and related B-type) asteroids are considered to be the most primitive bodies in the inner solar system as they are inferred to be relatively rich in volatile materials like water and organics. The only C-type asteroid visited by a spacecraft so far is the main belt asteroid (253) Mathilde, but it was a flyby observation and full global mapping images were not obtained (Thomas et al., 1999). Thus, we do not have yet any detailed information on the most primitive types of asteroids, other than laboratory analysis of primitive meteorites, but the specific provenance of these is unknown.

Hayabusa2 will arrive at Ryugu in the early summer (mid-June to beginning of July) of 2018 and stay around Ryugu for one year and a half (Tsuda et al. 2013; Watanabe et al. 2017). In addition, the NASA asteroid sample return mission OSIRIS-REx will arrive at a B-type asteroid (101955) Bennu (~500 m in diameter) in August of the same year (Lauretta et al. 2017). These will be the first volatile-rich asteroids for which we will obtain high-resolution images of nearly the whole surfaces and will provide critical insight as to whether observed characteristics of boulders, craters, and surface roughness (in addition to other properties discussed elsewhere in this paper) are representative of similarly sized, primitive asteroids.

Since Ryugu has never been visited before, the limited information before Hayabusa2's arrival should be reviewed and appropriately set out in order to guide the Hayabusa2 mission toward success. For this purpose, the International Regolith Science Group (IRSG) was organized as a branch of the Interdisciplinary Science Team in the Hayabusa2 project. In this paper, a product of an IRSG activity, we review the information about Ryugu before the encounter of Hayabusa2. We focus especially on the properties of the surface regolith layer, i.e. its global properties including surface geological features (Section 2), its thermophysical properties (Section 3), and its mechanical properties (Section 4). The information described in each section falls into three categories in terms of our present knowledge: the measured data obtained so far, the properties directly derived/deduced from the present data with basic physics, and the predicted properties. As a whole, this paper provides a reference model of Ryugu's regolith with the categorized data/properties. This model will be useful for Hayabusa2 operations such as landing site selection and touch-down operations as well as for interpreting and understanding data that will be obtained by the remote sensing observations of Hayabusa2 and by the analysis of samples returned from the asteroid.

Hayabusa2 carries a suite of instruments for scientific observation. These include three optical navigation cameras (ONC), consisting of a telescopic camera with seven band filters (ONC-T) and two wide field of view cameras (ONC-W1 and -W2), a near-infrared spectrometer (NIRS3), a thermal infrared imager (TIR), and a light detection and ranging (LIDAR) instrument. The surface of Ryugu will be investigated with these instruments. For example, global and local mapping of topography, such as craters and boulders, will be carried out by ONC imaging (Kameda et al. 2017). The spectroscopic properties of the surface will be revealed by ONC-T and NIRS3 observations, e.g., for detection of hydrated minerals (Kameda et al. 2015; Iwata et al. 2017). The thermophysical properties such as the thermal inertia derived from the surface temperature distribution will be explored by TIR observations, and these will help to estimate the particle size and the porosity of the regolith layer (Okada et al. 2017; Arai et al. 2017; Takita et al. 2017). LIDAR will be used for the albedo observation and dust detection around Ryugu, in addition to measuring the distance between the Hayabusa2 spacecraft and Ryugu's surface (Namiki et al. 2014; Mizuno et al. 2017; Yamada et al. 2017; Senshu et al. 2017). Two landers (rovers) are also carried by Hayabusa2, called MINERVA-II and MASCOT. In particular, MASCOT, developed by the German Aerospace Centre (DLR) in collaboration with the Centre National d'Etudes Spatiales (CNES), has several instruments for scientific investigation on the surface (Ho et al. 2017): a camera with illumination unit (MasCam, Jaumann et al. 2017), a near- to mid-infrared spectromicroscope (MicrOmega, Bibring et al. 2017), a multi-channel radiometer (MARA, Grott et al. 2017), and a magnetometer (MasMag, Herčík et al. 2017). The close observation of the regolith surface by each of MASCOT's instruments will provide "ground truth" of the remote sensing data obtained by the Hayabusa2 spacecraft, such as the particle size distribution, mineral composition, and thermal inertia. In addition to the above remote sensing data, the mechanical properties of the regolith layer will be investigated by an impact experiment carried out with the so-called Small Carry-on Impactor (SCI) and in-situ observation with a deployable camera (DCAM3) (Saiki et al. 2017; Arakawa et al. 2017; Ogawa et al. 2017; Ishibashi et al. 2017; Sawada et al. 2017). If the trajectory of MASCOT during its descent and after rebound on the surface is traceable, its analysis can also be used to estimate the mechanical properties of the regolith layer (e.g., Thuillet et al. 2018). After the end of the mission at Ryugu, and with the benefit of analyses of the returned samples, the reference model built in this paper will be fully checked and refined, and our knowledge of small bodies in the solar system will be considerably improved.

## 2. Global Properties

Here, the global properties of Ryugu are presented along with the best knowledge of the uncertainties in the measurements. The described properties provide essential inputs for mission planning in advance of Ryugu's arrival. In particular, the design reference mission requires precise knowledge of Ryugu's orbit and dynamical state, size and shape, and near-surface spectral reflectance properties.

### 2.1. Measured Quantities

#### 2.1.1. Orbital Properties

The observations used for orbit determination of Ryugu date from 1986, and include 725 measurements from the JPL-Horizons ephemerides (from 16-May-2017). The derived orbit is listed in Table 1. Ryugu's perihelion (q) of 0.963 AU, and aphelion (Q) of 1.416 AU classify it as an Apollo-type near-Earth object. Its Earth Minimum Orbit Intersection Distance (MOID) of 0.00111549 AU is among the smallest values for asteroids larger than 100 m (JPL/SSD).

Numerical modeling of NEA orbital evolution indicates that Ryugu originated in the inner asteroid belt. According to a model by Bottke et al. (2002), Ryugu is statistically most likely to have followed the $\nu_6$ secular resonance pathway from the Main Asteroid Belt (Campins et al. 2013). This pathway marks the inner boundary of the Main Asteroid Belt at ~2.1 AU and is the most efficient way to deliver bodies into near-Earth space, accounting for most of those that reach very Earth-like orbits characterized by low-MOID values and low Delta-V space mission trajectories. Several models suggest that Ryugu likely originated in the inner Main Asteroid Belt, between ~2.1-2.5 AU, and reached $\nu_6$ by inward Yarkovsky drift (Campins et al. 2013, Walsh et al. 2013, Bottke et al. 2015).

Working from the scenario of an inner Main Belt origin for Ryugu, it is possible to seek relationships with known asteroid families. Campins et al. (2013) suggested that the spectral taxonomy of Ryugu prevents a solid match with known families, while Bottke et al. (2015) studied the dynamical possibilities of delivery from either the New Polana or Eulalia families. This work found that Ryugu was more likely to have originated from the ~1-Gyr old New Polana family that is centered at 2.42 AU. Surveys of these two families in visible and near-infrared wavelengths find no significant differences between them, leaving Ryugu's origin uncertain (Pinilla-Alonso et al. 2016, de León et al. 2016).

**Table 1: Summary of important orbital and global properties. The orbital data are taken from JPL's SSD website (NASA's Jet Propulsion Laboratory Solar Systems Dynamics Group - HORIZONS at http://ssd.jpl.nasa.gov/). Rotation period is from numerous references (Kim et al. 2013; Müller et al. 2017; Perna et al. 2017).**

| | |
|---|---|
| Eccentricity | 0.190208 |
| Semi-major axis (AU) | 1.189555 |
| Inclination (deg) | 5.883556 |
| Period (days) | 473.8878 |
| Perihelion (AU) | 0.963292 |
| Aphelion (AU) | 1.415819 |
| Rotation Period (hr) | 7.6326 |

### 2.1.2. Size, Shape, and Spin

The reference Ryugu shape model was reconstructed from the inversion of optical and thermal infrared data by Müller et al. (2017). Because of the small amplitude of the obtained light curves and low quality of the optical light curves, the inversion did not lead to a unique solution for the pole, shape, and period. Formally, there were different sets of these parameters that fit the available data equally well. However, after a careful analysis, Müller et al. (2017) concluded that the most likely pole direction in ecliptic coordinates is: lambda = 310-340°, beta = -40° ± 15°. The corresponding shape model is roughly spherical with a volume-equivalent diameter of 850-880 m (see Figure 1).

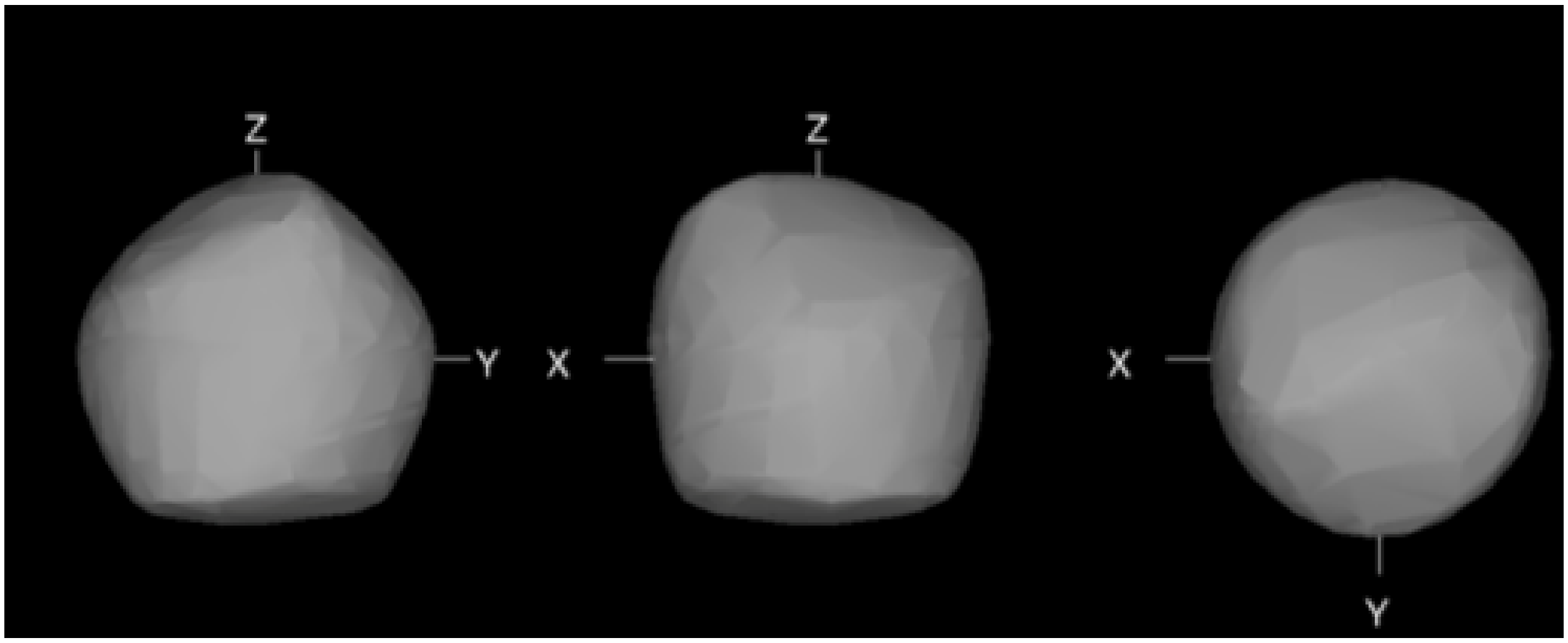

Figure 1: **Best fit shape model for Ryugu. Image taken from Müller et al. 2017.**

Ryugu has a rotation period of approximately 7.63 h (Perna et al. 2017). Müller at al. (2017) provided more possible values for the sidereal rotation period (7.6300, 7.6311, 7.6326 h) but from the analysis of the new data from Very Large Telescope (VLT) (Perna et al. 2017) and other observatories (Kim. et al. 2016), the sidereal rotation period of 7.6326 h is preferred (Durech, pers. comm.). As shown in Figure 2, the light curve has an asymmetric behavior with two different maxima and minima. The asymmetric trend of the light curve is connected to either the irregular shape and/or variations in albedo.

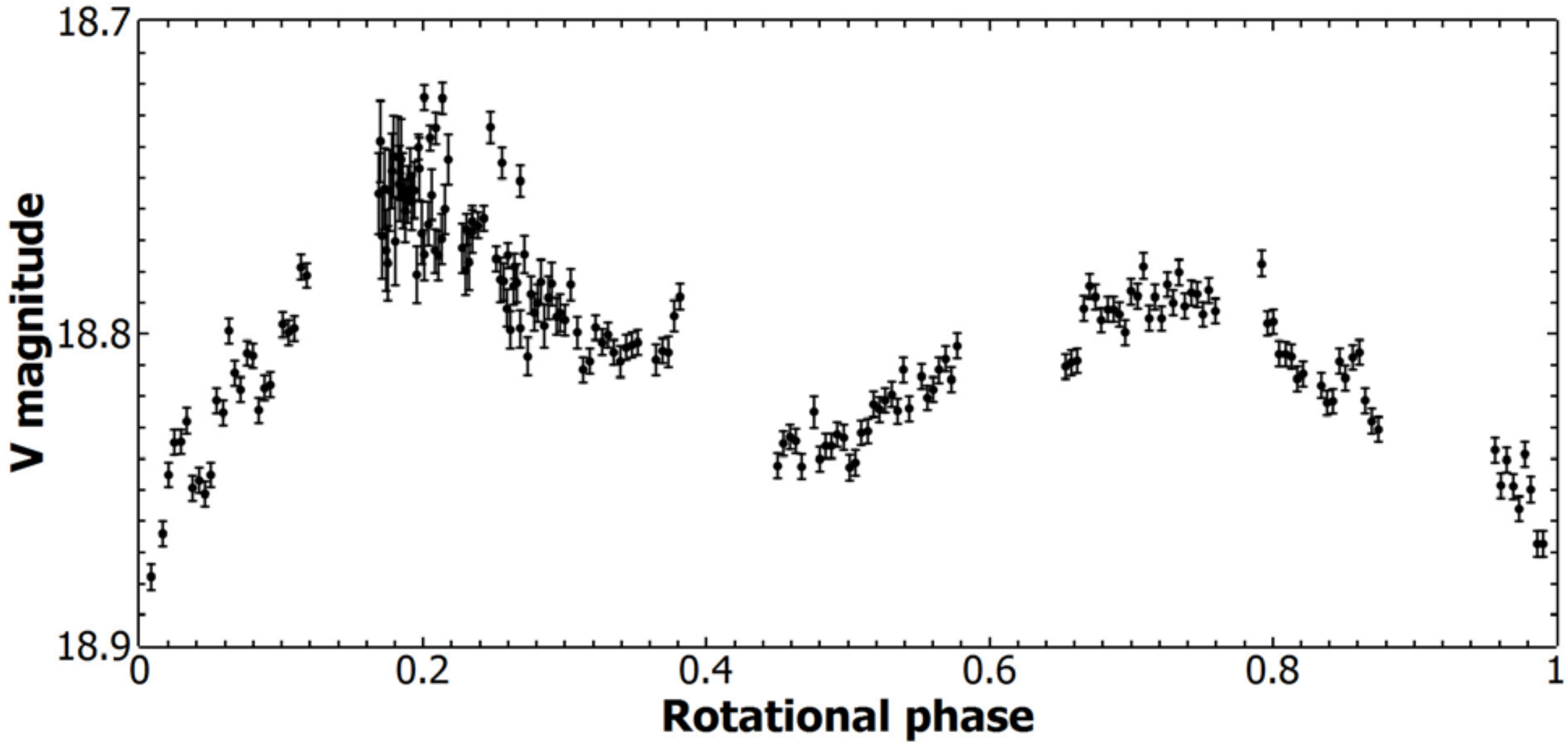

**Figure 2: Composite lightcurve of Ryugu, as observed at the Very Large Telescope of the European Southern Observatory (ESO-VLT) on 12-Jul-2016 by Perna et al (2017). The rotational phase corresponds to a synodic period of 7.63 h. The zero phase time corresponds to 57581.0 MJD.**

### 2.1.3. Phase Function

As discussed in Le Corre et al. (2018), the spherical albedo and phase integral of Ryugu are useful quantities for deriving the bolometric Bond albedo map. These quantities allow us to make temperature prediction across the surface of this asteroid during site and landing selections events of Hayabusa2 mission. These quantities can also be useful to calculate Ryugu's thermal inertia and constrain the Yarkovsky effect.

Le Corre et al. (2018) used ground-based photometric data of Ryugu, which were compiled by Ishiguro et al. (2014) and references therein, to constrain the average disk-resolved brightness across Ryugu's surface. The Radiance Factor (*RADF*) is the ratio of the bidirectional reflectance of a surface to that of a perfectly diffuse surface illuminated at the incident angle $i$ = 0 (Hapke 2012). Reflectance, $r_{ls}(i,e,\alpha)$, is directly related to $RADF(i,e,\alpha)$ or $[I/\mathcal{F}](i,e,\alpha)$ as described in:

$$[I/\mathcal{F}]\ (i,e,\alpha) = RADF(i,e,\alpha) = \pi r_{ls}(i,e,\alpha) \qquad (1),$$

where $i$, $e$, and $\alpha$ being the incidence angle, the emission angle, and the phase angle, respectively. $I$ is the radiance and has units of W m$^{-2}$ nm$^{-1}$ sr$^{-1}$. $J = \pi\mathcal{F}$ is the irradiance and has units of W m$^{-2}$ nm$^{-1}$.

Under the assumption that the photometric properties of the surface are well described by a Lommel-Seeliger law, we can write:

$$[I/\mathcal{F}]\ (i,e,\alpha) = \frac{w_o}{4}\ \frac{\mu o\ f(\alpha)}{\mu o + \mu} \qquad (2),$$

where $\mu_o = \cos(i)$, $\mu = \cos(e)$, $A_{LS} = \frac{w_o}{4\pi}$ is the Lommel-Seeliger albedo, $f(\alpha)$ is the phase function, and $w_o$ is the average particle single scattering albedo. We approximate the phase function with an exponential-polynomial function of the form $f(\alpha) = e^{\beta\alpha+\gamma\alpha^2+\delta\alpha^3}$ (Takir et al. 2015), the coefficients of which, $\beta$, $\gamma$, and $\delta$, are fitted to the data.

Figure 3 shows Lommel-Seeliger models that capture low and high phase angle behavior, and the scatter in the moderate phase angle ground-based observations of Ishiguro et al. (2014). Le Corre et al. (2018) computed models for nominal, maximum, and minimum predicted brightness of Ryugu at 550 nm (Table 2). Although this photometric model fitted the disk-integrated data of Ryugu well, It may not be an appropriate model for this asteroid's disk-resolved data especially at larger phase angles (>100$^{o}$) (e.g., Schröder et al. 2017). This Lommel-Seeliger model is a preliminary model and can be updated when the spacecraft arrives at Ryugu.

Table 3 includes the albedo quantities of Ryugu, computed with the Lommel-Seeliger model, and the quantities computed by Ishiguro et al. (2014). The two sets of quantities are consistent with each other.

**Table 2: Lommel-Seeliger functions that predict $[I/\mathcal{F}]$ ($i$,$e$,$\alpha$) (reflectance) of Ryugu at 550 nm. $A_{LS}$ is Lommel-Seeliger Albedo and $f(\alpha) = e^{\beta\alpha+\gamma\alpha^2+\delta\alpha^3}$. Table from Le Corre et al. (2018).**

| | $A_{LS}$ | β | γ | δ |
|---|---|---|---|---|
| Nominal | 0.027 | -4.14 x $10^{-2}$ | 3.22 x $10^{-4}$ | 19.16 x $10^{-7}$ |
| Maximum | 0.030 | -3.99 x $10^{-2}$ | 3.27 x $10^{-4}$ | 22.15 x $10^{-7}$ |
| Minimum | 0.024 | -4.46 x $10^{-2}$ | 3.77 x $10^{-4}$ | 21.33 x $10^{-7}$ |

**Table 3: Albedo quantities of Ryugu computed using the Lommel Seeliger model and by Ishiguro et al. (2014). $p_v$ is the geometric albedo, $q$ is the phase integral, and $A_B$ is the spherical Bond albedo. Table from Le Corre et al. (2018).**

| | $p_v$ | $q$ | $A_B$ |
|---|---|---|---|
| Lommel-Seeliger | $0.042^{+0.005}_{-0.004}$ | $0.34^{+0.01}_{-0.01}$ | $0.014^{+0.001}_{-0.001}$ |
| Ishiguro et al. (2014) | $0.047^{+0.003}_{-0.003}$ | $0.32^{+0.03}_{-0.03}$ | $0.014^{+0.002}_{-0.002}$ |

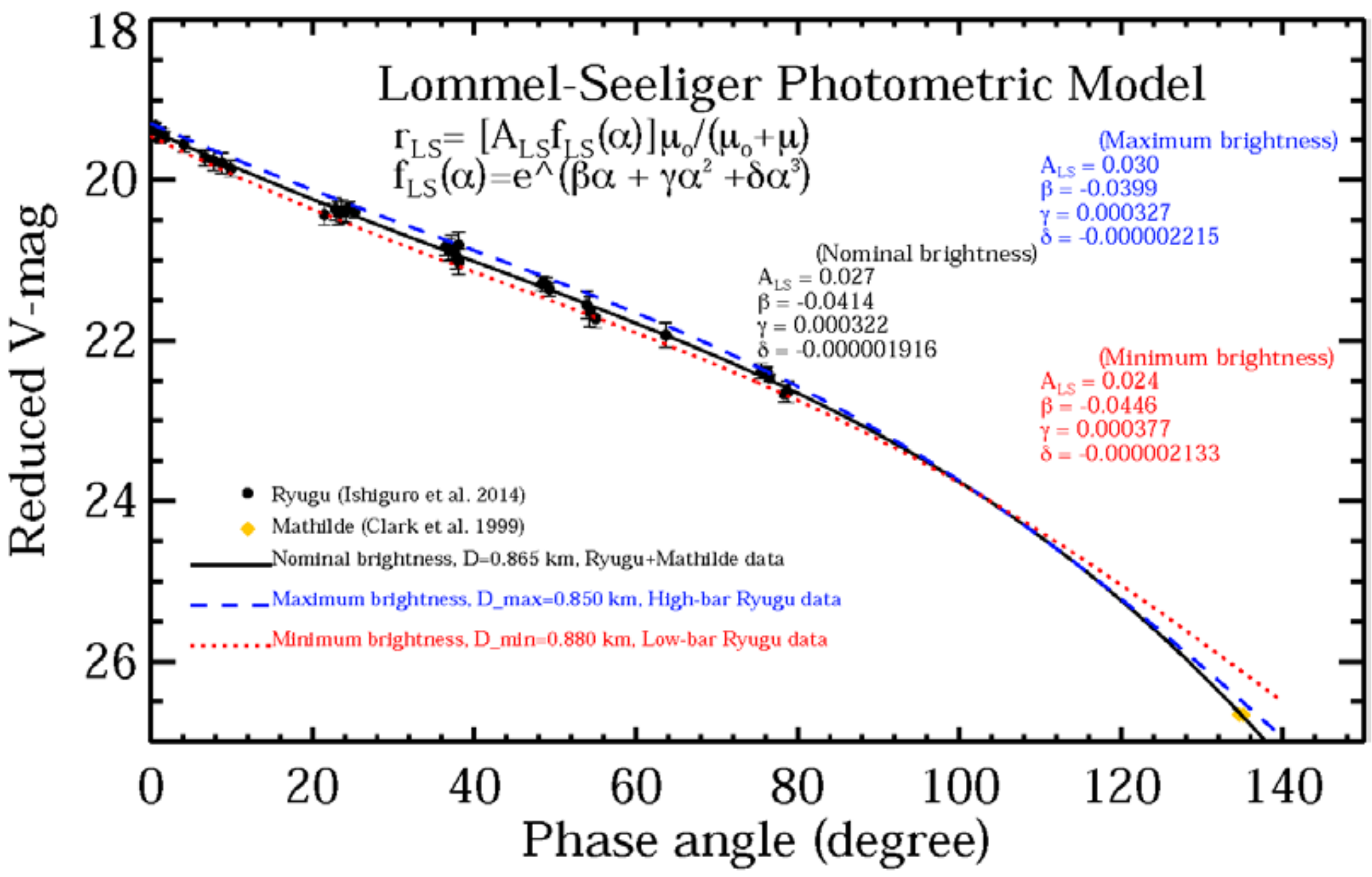

**Figure 3: The Reduced V magnitude of Ryugu as a function of phase angle predicted by the Lommel-Seeliger model is shown compared with the ground-based measurements of Ishiguro et al. (2014) and references therein. Shown are the minimum**

**(red dots), maximum (blue dashes), and nominal (black solid line) models. Our minimum and maximum models do not include the Mathilde data, however our nominal model does. Reflectance $r_{ls}$ is in units of sr$^{-1}$. Figure from Le Corre et al. (2018).**

### 2.1.4. Spectral Properties

Ryugu is a primitive Apollo NEA classified as a C-type asteroid. At present, there are 8 ground-based spectroscopic observations of this asteroid, three of which extends towards the near-IR (Table 4). In nearly all the observations, the spectrum is featureless, with a slight reddening in the IR and a slight UV drop off. The best meteorite spectral counterpart is a heated CM chondrite or possibly a CI chondrite (Tonui et al. 2014), however this comparison is complicated by the poorly-known effects of space weathering on Ryugu.

The spectral analysis by Vilas (2008) at visible wavelengths showed a feature at 0.7 µm attributed to the $Fe^{2+}$->$Fe^{3+}$ charge transfer transition in oxidized phyllosilicates. This feature has not been detected in any other observations of Ryugu and the author suggested a spatial heterogeneity in the asteroid surface composition, as already observed in other main-belt C-type asteroid spectra (see Rivkin et al. 2002).

The observations by Lazzaro et al. (2013) in the 0.4-0.85 µm spectral range covered 70% of Ryugu's surface and showed an almost uniform C-type featureless spectrum, with no direct evidence of hydration features. Few variations have been observed at wavelengths shorter than 0.5 µm, which lie in a UV drop-off as found in previous observations (Binzel et al 2001). The UV drop-off is a common spectral feature of asteroids whose taxonomy is associated with carbonaceous chondrites, particularly for G and F asteroid classes (Tholen & Barucci 1989) or Cg in the Bus-DeMeo system (Bus and Binzel 2002; DeMeo et al. 2009). This spectral behavior was investigated and widely discussed by Hiroi et al. (1996, 2003) and Cloutis et al. (2012). These authors noticed decreased reflectance in the UV after heating or laser bombardment processing to simulate micrometeorite bombardment. These hypervelocity impacts cause localized melting and the production of spherical glassy droplets. The slight UV drop-off observed in different regions of Ryugu's surface could thus be explained by the effect of heterogeneous processing due to space weathering and/or episodes of substantial heating via, presumably, closer perihelion passages.

The IR observations by Abe et al. (2008) and Pinilla-Alonso et al. (2013), show a featureless/flat spectrum and confirm that Ryugu is a C-type asteroid in line with all the other observations. A very slight positive slope is observed. The two studies derive different slope values, which could be indicative of a heterogeneous surface; however, the visible observations by Moskovitz et al. (2013) reported a homogenous and flat spectrum in different observations of Ryugu, suggesting that possible heterogeneities should be constrained to 5% of the entire asteroid surface. This work proposed that Ryugu was best represented by a thermally altered sample of the Murchison (CM2) meteorite and by the thermally metamorphosed CI chondrite Yamato 86029. The most recent work by Perna et al. (2017) confirms all the previous observations with a slight UV drop off and a featureless/flat spectrum in both the visible and near IR (Figure 4). Le Corre et al. (2018) observed Ryugu during its close fly-by of the Earth in July 2016. Their spectrum differs from those presented in Moskovitz et al. (2013) and Perna et al. (2017), with a stronger red spectral slope shortward of 1.6 µm. However, like previous observations, Ryugu's spectrum does not show any well-defined absorption bands. Le Corre et al. (2018) suggested the possible presence of two broad absorption bands centered around 1 and 2.2 µm in their data. They proposed that if both are present, these bands could be indicative of a CO or CV chondrite-like composition. These meteorites are typically too bright to match Ryugu's reflectance and would require darkening effects of space weathering to provide a match to the asteroid (Le Corre et al. 2018). These authors used curve fitting techniques to confirm that CM chondrites remain the best match. They also found two small asteroids with very similar spectra (showing a pronounced red slope) to Ryugu: NEA (85275) 1994 LY and Mars-crossing asteroid (316720) 1998 BE7. In addition, this work showed that the newly-observed spectrum of Main Belt asteroid (302) Clarissa, suggested as a possible source family for Ryugu (Campins et al., 2013), does not match their spectrum of Ryugu. It is however a closer match with the spectrum from Moskovitz et al. (2013).

**Table 4: Spectroscopic observations of asteroid Ryugu since its discovery in 2001. Spectral and compositional characteristics are listed.**

| Reference | Spectral interval µm | Taxonomy Type T=Tholen BD=Bus-De Meo | Meteoritical counterpart | Spectral Characteristics |
|---|---|---|---|---|
| Binzel et al. 2001 | 0.4-0.9 | Cg (BD) | N.A. | UV drop off: Red (0.4-0.65) µm Flat (0.65-0.9)µm |
| Vilas 2008 | 0.42-0.93 | C (T) | CM2 | Band at 0.7 µm Flat with very weak band at 0.6 µm |
| Abe et al. 2008 | 0.4-2.4 | C | N.A. | Flat & Featureless spectral slope (0.85-2.2 µm) 0.89 ± 0.03 %/1000 Å |
| Lazzaro et al. 2013 | 0.4-0.85 | C | N.A. | Flat & Featureless UV drop off (<0.45 µm) |
| Pinilla-Alonso et al. 2013 | 0.85– 2.2 | C | N.A. | Featureless spectral slope (0.85-2.2 µm) 0.37 ± 0.28 %/1000 Å |
| Sugita et al. 2013 | 0.47-0.8 | 163Erigone Family | CM (heated Murchison) | Flat & Featureless |
| Moskovitz et al. 2013 | 0.44-0.94 | C (BD) | CM (heated Murchison) CI (Y-86029) | Spectrally flat |
| Perna et al. 2017 | 0.35-2.15 | C (BD) | heated Murchison CM unusual/heated CI | Featureless UV drop off (<0.45 µm) spectral slope (0.5-0.8 µm) |
| Le Corre et al. 2018 | 0.8-2.4 | C | CM2 carbonaceous chondrites, Mighei (grain size < 40 µm) and ALH83100 (grain size < 100 µm) | near-perfect match with NEA (85275) 1994 LY and Mars-crossing asteroid (316720) 1998 BE7 |

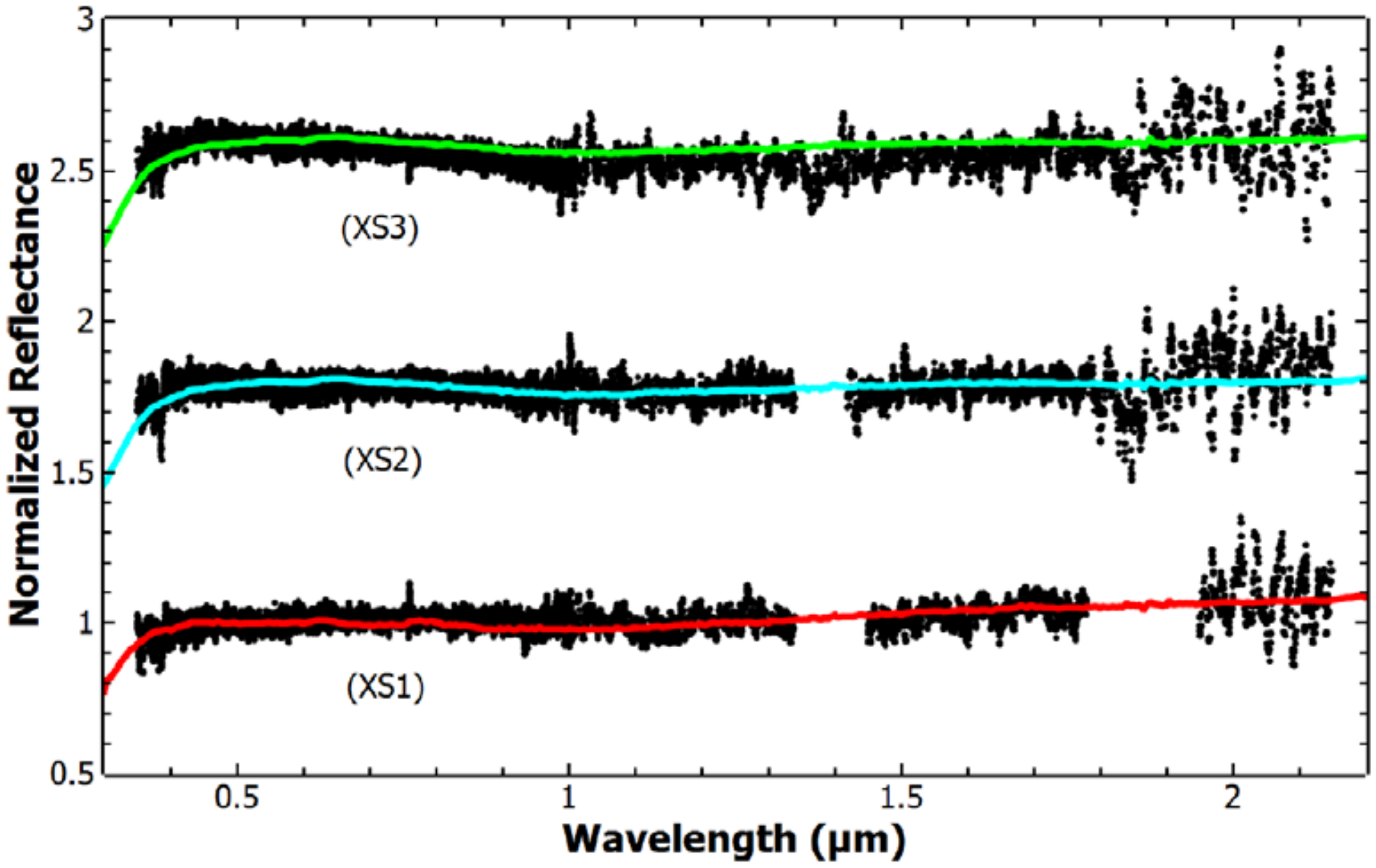

**Figure 4: The featureless/flatness nature of the Ryugu spectrum is evident in this series of three Vis-IR spectra. Superimposed are the spectra of thermally altered samples of the CM carbonaceous chondrite Murchison, in red (heated at 900 °C) and in cyan and green (heated at 1000 °C) (from Perna et al. 2017).**

## 2.2. Derived Quantities

Determining the composition of Ryugu by remote spectroscopic observations requires accounting for viewing geometry and modeling of the asteroid's surface properties. Reflectance properties of asteroids may differ from laboratory analogues due to grain size, phase angle, and space weathering (Binzel et al. 2015; Sanchez et al. 2012; Reddy et al. 2012). Laboratory spectral measurements of meteorites relevant to Ryugu can improve the interpretation of future hyperspectral observations obtained by Hayabusa2 NIRS3 instrument by linking absorption bands to well-characterized mineralogical composition. Here, we summarize various authors' interpretations of their Ryugu spectra, most of which are included in Table 4 (see Le Corre et al. 2018 for a lengthy discussion of spectral interpretations of Ryugu).

### 2.2.5. Composition

Generally, ground-based observations of asteroid Ryugu show a featureless spectrum, with a slight reddening in the near IR and with a slight drop off in the UV range. Based on laboratory data, the best available meteoritic spectral analogs are, to first order, a heated CM or possibly CI chondrites. However, a list of possible surface analogs for Ryugu should contain at least components of CM and CI meteorites (e.g. Murchison, Orgueil, Jbilet Winselwan, Y82182, Y86029 and so on), plus a suite of serpentines, saponite, phyllosilicates, magnetite, sulfates, sulfides, pyroxenes, and carbonates.

The extent of alteration and the spectral characteristics of 16 heated CI and CM chondrites was investigated at the Planetary Science Laboratory (PSL) of the Institute for Planetary Research of DLR, in Berlin, Germany. Only meteorites whose bulk modal mineralogy and $H_2O$ contents had previously been examined were included. Near- and mid-IR reflectance spectra were collected for powdered samples of the meteorites. The aim was to directly relate spectral features to the known properties and alteration history of the heated CI and CM chondrites and accurately interpret the surface mineralogy of C-type asteroids. Features in the mid-IR are particularly diagnostic of the anhydrous and hydrous silicate mineralogy and can be used to remotely infer the extent of aqueous and thermal processing on these primitive bodies (Tonui et al. 2014; King et al. 2016).

Laboratory measurements of reflectance spectra of a series of 23 carbonaceous chondrites belonging to the CM and CR groups have been obtained between 0.3 to 100 µm in support of efforts to be able to interpret and relate measurements of the surface of Ryugu. The major results are i) the presence of 0.7 and 0.9 µm features correlated to the amount of phyllosilicates (confirming previous work such as Vilas and Sykes 1996); ii) reporting for the first time of the presence of 0.7 and 0.9 µm features in spectra of a CR chondrites (Grosvenor Mountains 95577); these data suggest that Cgh-type asteroids (which make up about 1/3 of the main-belt C-type asteroids) could be the parent bodies of CM but also of some CR chondrites; iii) the presence of a goethite-like 3-µm band for some CR chondrites (Beck et al. 2018).

## 2.3. Predicted Quantities

Moving from derived quantities, where measured properties are used to derive new values for the properties of Ryugu, here we use the best evidence regarding Ryugu to predict properties for which there are no measurements of any kind. This section will often rely on knowledge gained from previous explorations and general estimates for C-type and NEAs. Some are explained in depth, notably block sizes, due to the importance they could have for mission operations planning regarding spacecraft safety.

### 2.3.6. Satellites, Dust and the Local Environment

There are no directly observed, or inferred, satellites at Ryugu, so this feature, along with dust floating around the asteroid, has fallen into the “predicted” category of this work. What would drive a prediction of the existence of a satellite? Satellites among NEAs are common – it is expected that 15% of the population have satellites on relatively close orbits (Walsh and Jacobson 2015, Margot et al. 2015). However, nearly all NEAs with satellites are very rapidly rotating, typically near the critical rotation rate of ~2.2h, and nearly all have small lightcurve amplitudes, and where available, their shapes are often found to be "top-shapes" or nearly spherical with equatorial bulges. Given that small bodies can be spun up or down by the Yarkovsky–O'Keefe–Radzievskii–Paddack effect (YORP), but this is highly uncertain, the current spin rate of Ryugu certainly does not indicate that it was ever a rapid rotator in the past, but at first glance there is no strong indication that Ryugu will have a satellite or significant floating dust.

### 2.3.7. Boulders, Craters, and Surface Roughness

Safely delivering the spacecraft to the asteroid surface and effectively obtaining a sample are mission critical events that require careful evaluation of the asteroid surface properties. The distribution and abundance of boulders and craters pose the greatest risk to spacecraft flight that must be carried out autonomously. The surface roughness, or regolith grain size distribution are irrelevant to safety but critical to the ability to successfully obtain a sample. Here we summarize spacecraft observations of asteroids that provide useful guidance for what we can expect upon arrival to asteroid Ryugu.

Asteroid Itokawa is similar in size (535 m x 294 m x 209 m; Demura et al. 2006) to Ryugu (~ 900 m), but significant differences in geological features are expected. Returned sample analyses confirmed that S-type asteroid Itokawa is composed of ordinary chondrite-like material, whereas C-type asteroid Ryugu is likely to be more volatile rich and weaker. It is already known that Itokawa and Ryugu have strikingly differing shapes. Itokawa is highly elongated and appears to be a rubble pile composed of fragments from the catastrophic disruption of a larger body (Michel and Richardson 2013). In contrast, Ryugu is roughly spherical, but its evolutionary history is not yet known.

The only C-type asteroid that has been previously visited by spacecraft is the main belt asteroid Mathilde—a significantly larger asteroid at ~53-km mean diameter (Thomas et al. 1999). The best images obtained during the NEAR flyby of Mathilde were 160 m/pixel (Thomas et al. 1999). At this scale, Ryugu would be only a few pixels across, so these images do not offer much insight into what to expect on Ryugu's surface. Several other asteroids (S-types Eros, (243) Ida, (951) Gaspra, (2867) Steins, (4179) Toutatis, (5535) Annefrank; Q-type, (9969) Braille; M-type (21) Lutetia) and other small bodies (Phobos, Deimos, several comets, and moons of Saturn) have been imaged by spacecraft and provide some constraints as to what to expect from Ryugu; however, of those bodies only Eros, Phobos, Toutatis, and 67P/Churyumov-Gerasimenko (67P) were imaged at scales sufficient to resolve small boulders, and only Eros, Phobos (highest resolution images only available in one area), and 67P have been imaged globally. The size of Eros, the size and celestial location of Phobos, and the cometary composition/nature of 67P all pose complications for using these bodies as analogs to Ryugu.

At a larger scale, images of Ida and Gaspra look broadly similar to those of Eros, perhaps indicating that Eros is representative of an S-type asteroid of its size. Aside from Itokawa, Steins (5.3-km mean diameter) and Toutatis (3-km mean diameter) are the asteroids that have been observed by spacecraft that are closest in size (both are larger) to Ryugu but have a different taxonomic type; the best global Steins images have a ~160-m pixel scale, again insufficient to give much insight into Ryugu. The Toutatis data are limited, but have a pixel scale of ~3 m. There is limited, very low resolution data available for Annefrank and Braille. (4) Vesta and Ceres are so much larger than Ryugu that they do not provide very good points of comparison for what to expect.

### 2.3.8. Size Frequency Distribution of Boulders

The cumulative size–frequency distributions (SFD) of boulders on many small bodies (where sufficient data are available) can be fit by power laws (see Table 5). Many, though not all, of the fits to global boulder populations have power-law exponents around -3. The exponent of the power-law fit depends on several factors, including the geological context, the strength of the material, and possibly the size of boulders. Power-law behavior of cumulative size–frequency distributions of terrestrial fragmented objects has also been observed (Turcotte 1997; see also Table 1 in Pajola et al. 2015), with the examples listed in Pajola et al. (2015) having exponents ranging from -1.89 to -3.54.

In the cases of Eros, Itokawa, and Phobos, extending the SFD power-law fit from large, tens-of-meter-sized blocks down to small, tens-of-centimeter-sized blocks yields reasonable estimates of small block populations (Rodgers et al. 2016, see Figure 5). However, the geological context of an area matters for the absolute block density – if lower-resolution counts include multiple geological settings, they will not extrapolate accurately to local areas containing only one setting (Ernst et al. 2015; Rodgers et al. 2016).

**Table 5: Small bodies for which boulder counts have been made from spacecraft images. The minimum boulder sizes measured are directly related to the best image resolution available for a given object.**

| Name | Mean Diameter (km) | Spectral Type | Min boulder size of global count (m) | Min boulder size of regional count (m) | Power law found | Data source | References |
|---|---|---|---|---|---|---|---|
| Eros | 17 | S | 15 | 0.05 | -3.2 as low as -2.3 locally | NEAR | Rodgers et al. 2016; Thomas et al. 2001 |
| Itokawa | 0.35 | S | 6 | 0.1 | -3.1<br>-3.5<br>as low as -2.2 locally | Hayabusa | Mazrouei et al. 2014; Michikami et al. 2008; Noviello et al. 2014; Rodgers et al. 2016 |
| Toutatis | 2.9 | S | n/a | 10 | -4.4 locally | Chang'e 2 | Jiang et al. 2015; Huang et al. 2013 |
| Lutetia | 99 | M | n/a | 60 | -5.0 | Rosetta | Küppers et al. 2012; Sierks et al. 2011 |
| Ida | 32 | S | n/a | 45 | n/a | Galileo | Lee et al. 1996 |
| Phobos | 22 | D | n/a | 5 | -3.2 | Viking MGS<br>MEX<br>MRO | Ernst et al. 2015; Rodgers et al. 2016; Thomas et al. 2000 |
| Deimos | 12 | D | n/a | ~4 | -3.2 | Viking | S. W. Lee et al. 1986;<br>C. Ernst, personal communication |
| Churyumov-Gerasimenko | 4 | comet | 7 | 2 | -3.6 global<br>local ranges -2.2 to -4.0 | Rosetta | Pajola et al. 2015; 2016 |

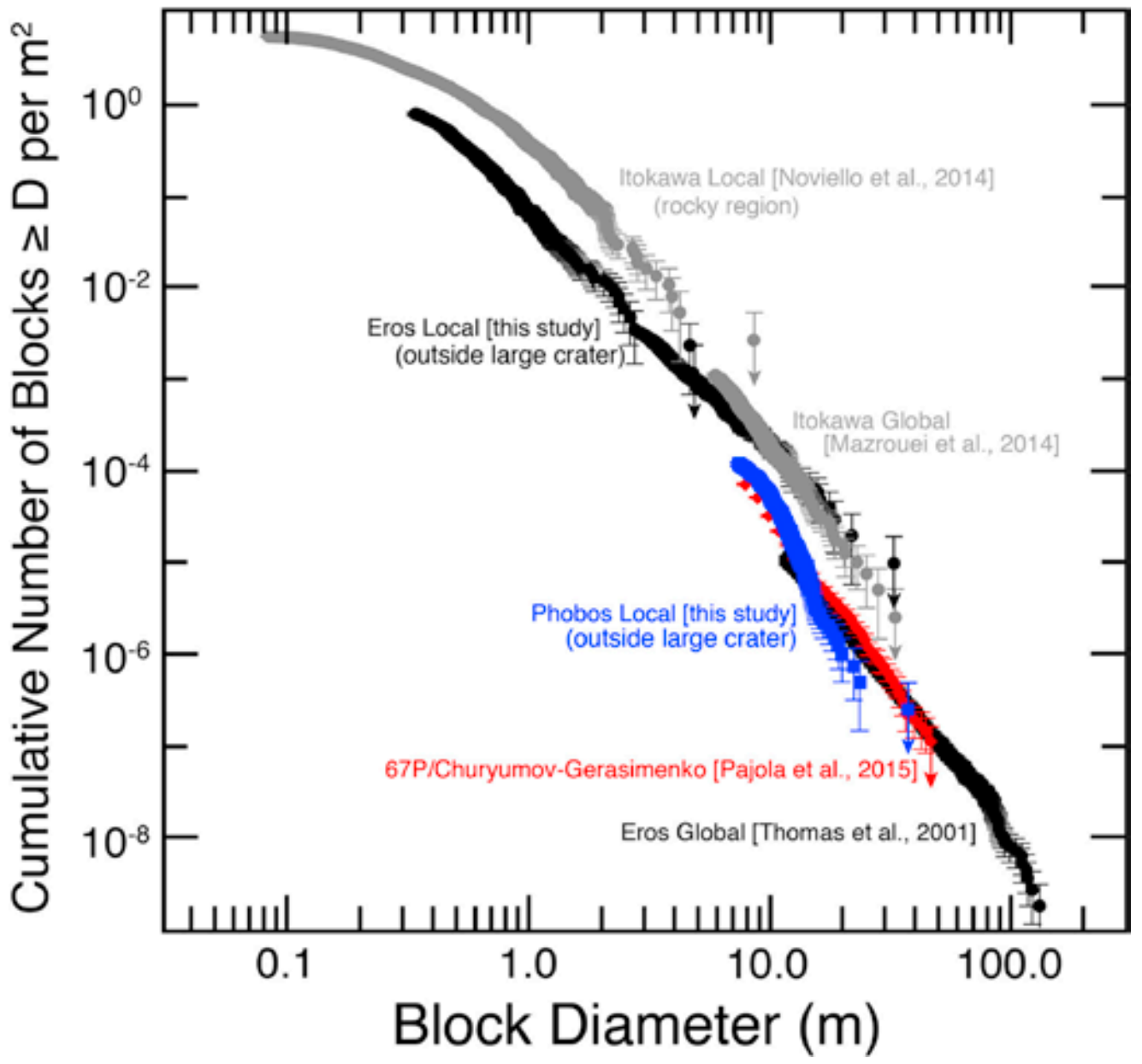

**Figure 5: Figure taken from Rodgers et al. 2016 showing the size frequency distribution of the blocks measured on Eros, Itokawa, Phobos and comet 67P/Churyumov-Gerasimenko (see Rodgers et al. 2016, Figure 1.).**

Radar observations of some NEAs have identified large boulders on at least 14 NEAs (Benner et al. 2015), but the size of the boulders identified is limited by the resolution of the data. Toutatis provides the only "ground truth" between radar identifications of boulders and spacecraft images of boulders—the SNR of the radar observations was too low to positively identify boulders on Eros or Itokawa (Benner et al. 2015). Radar data can be used to measure the radar scattering properties of asteroids, which gives insight into surface cobbles of the scale of the radar wavelength (0.035 m for Goldstone, 0.126m for Arecibo).

If there were radar data that suggest either a number of boulders or a cobble number, an average size distribution of boulders for the asteroid could be estimated by assuming a -3 power law exponent. The actual distribution of boulders will be influenced by factors such as their origin, and the geopotential of Ryugu. Ponds full of cm-sized cobbles are found on Itokawa at geopotential lows, and most boulders are located in other regions of the asteroid (Fujiwara et al. 2006; Saito et al. 2006). If this distribution is typical of a small asteroid, Ryugu should have concentrations of finer material (ponds) near the geopotential lows, and boulders will be found at higher geopotentials. Generally, the lowest potential for an asteroid of Ryugu's shape is expected to be found at or near the equator; the shape and density of the asteroid will affect the location of these geopotential lows.

If the boulders are sourced from craters, as with Eros (Thomas et al. 2001), they are likely to be distributed around the craters as ejecta would fall on the irregularly shaped body. On the other hand, they may preferentially settle in the equatorial or near-equatorial regions.

### 2.3.9. Shapes of Boulders

The aspect ratios of boulders on Itokawa >6-m in diameter were measured by Mazrouei et al. (2014), who found that most boulders in this size range were elongated, with *b/a* ratios of 0.7 (*a*, *b*, and *c* are triaxial dimensions of the boulders, defined to be $a \geq b \geq c$). It was not possible to measure the third dimension of most boulders in the Hayabusa AMICA images. Similar *b/a* ratios (0.62–0.68) were found by Michikami et al. (2010). The *c/a* ratio was measured for 21 boulders, the mean was 0.46 (Michikami et al. 2016). A compilation of fragment dimensions from several laboratory experiments is shown in Table 6. Although the measurements assume an ellipsoidal shape, the fragments are actually irregular in shape. It is not known whether the aspect ratio of rocks on a C-type asteroid would be different from those seen on Itokawa or in the strong-rock experiments of Table 6. Also unknown is the magnitude of the influence of thermal degradation on boulder fragmentation.

**Table 6: Compilation of fragment ratios *b/a* and *c/a* from several publications in the literature. Dimensions are defined to be $a \geq b \geq c$.**

| Reference | Target | Projectile | Impact Velocity (km/s) | *b/a* | *c/a* |
|---|---|---|---|---|---|
| Fujiwara et al. 1978 | Basalt | Polycarbonate cylinders | 2.6 and 3.7 | 0.73 | 0.50 |
| Capaccioni et al. 1986 | Basalt<br>Concrete | Aluminum spheres | 8.8 and 9.7 | 0.7 | 0.5 |
| Durda et al. 2015 | Basalt | Aluminum spheres | 3.9-5.8 | 0.72 ± 0.13 | 0.39 ± 0.13 |
| Michikami et al. 2016 | Basalt | Nylon spheres | 1.6–7.1 | 0.74 | 0.5 (catastrophic disruption)<br>0.2 (impact cratering) |

### Craters

Itokawa has a small number of identified impact craters, especially small ones (Hirata et al. 2009). Given its similar size to Itokawa, Ryugu may also not exhibit many craters. Eros, Phobos, Ida, and Gaspra all have numerous craters; on Eros, seismic shaking is suggested to have acted to erase many small (<100-m-diameter) craters (Chapman et al. 2002; Thomas and Robinson 2005); craters of 100 m-size would be of significant size compared to Ryugu. These observations imply that Eros has a global loose regolith as deep as 100 meters (Robinson et al. 2002). Eros's relatively flat topography (<55% of the surface is steeper than 30°) may also be explained by a thick global regolith (Zuber et al. 2000). Craters are more apparent on Toutatis than on Itokawa (Huang et al. 2013), with different densities of craters seen on the different lobes of the asteroid (Zou et al. 2014). As Ryugu is intermediate in size between Itokawa and Toutatis, it might be expected to have more craters per unit area than Itokawa but fewer than Toutatis. Since the size of regolith particles on Ryugu is expected to be similar to that on Itokawa (see Section 3.2.2), steep cratered topography is not expected on Ryugu as well as Itokawa. However, the different composition of Ryugu could affect the size of craters and the regolith depth and structure.

### *Meter-Scale Surface Roughness*

Here, surface roughness at the meter-scale is described, to be distinguished from the smaller-scale roughness that is a parameter in thermophysical model fitting routines, or roughness in terms of very fine scale regolith grains – both are discussed in later Sections. The surface roughness of Itokawa and Eros has been measured by rendezvous spacecraft laser altimeters. Itokawa's two major regions, the highlands and lowlands, have different surface roughnesses that are obvious even in the images. The roughest parts of the asteroid correspond to those with more boulders, which correspond to regions of higher geopotential (Barnouin-Jha et al., 2008). It is thought that finer materials may move to regions of lower geopotential and cover up larger boulders, creating a smoother (less rough) region. These areas are thought to correspond to areas with the thickest (fine) regolith. Over baselines of 8–50 meters, the highlands of Itokawa are 2.3 × 0.4 m rougher than the MUSES-C regio (Barnouin-Jha et al. 2008); this difference may correspond to the expected regolith depth in this area.

The highlands of Itokawa are at a baseline of 20 meters - rougher than Eros - and the lowlands of Itokawa are significantly smoother than Eros (with the possible exception of Eros' ponds) (Barnouin-Jha et al. 2008). The surface roughness of Itokawa is not self-affine over the examined baselines (Barnouin-Jha et al. 2008)

On Eros, the surface roughness is self-affine over two orders of magnitude of baseline (~4–200 m), suggesting similar processes control the surface properties over these scales (Cheng et al. 2002). The corresponding Hurst exponents lie between 0.81 and 0.91 (Cheng et al. 2002). There are regional variations in surface roughness of Eros; smaller roughness values are found where more regolith is thought to have accumulated, and higher values are found along crater walls and grooves (Susorney and Barnouin 2016).

There is a strong dependence of radar circular polarization ratio with asteroid class, suggesting a relationship between roughness at the cm- to dm-scale and composition (Benner et al. 2008). However, the mean ratios for S- and C-type NEAs are indistinguishable, perhaps indicating similar average surface properties for these populations (Benner et al. 2008).

## 3. Regolith Thermophysical Properties

Regolith thermophysical properties govern the exchange of radiative energy between the asteroid and its environment, and knowledge of these parameters is needed to calculate surface and subsurface temperatures. While the surface energy balance is governed by insolation and regolith thermal inertia, heat diffusion governs temperatures in the subsurface. Heat is conducted to the subsurface according to

$$\rho c_p \frac{\partial T}{\partial t} = \frac{\partial}{\partial z} k \frac{\partial T}{\partial z} \tag{3}$$

where $\rho$ is regolith density, $c_p$ is heat capacity, $T$ is temperature, $t$ is time, $z$ is depth, and $k$ is thermal conductivity. Eq. (3) is a second order differential equation, which can be solved by prescribing two boundary conditions: One is usually given by zero heat flux from the interior, while the other is usually given in terms of the surface energy balance. For periodic insolation forcing, the surface energy balance takes the convenient form

$$\sigma_B \varepsilon T^4 = (1 - A)S + I\sqrt{\frac{\pi}{P}} \frac{\partial T}{\partial z'}\bigg|_{z'=0} \tag{4}$$

where $\sigma_B$ is the Stefan-Boltzmann constant, $\varepsilon$ is surface emissivity, $A$ is the Bond albedo, $S$ is total solar radiative flux including scattered radiation, $P$ is the period of the forcing, and $z' = z/d_e$ is depth normalized to the skin depth $d_e = \sqrt{kP/\varrho c_p \pi}$. In Eq. (4), all material parameters have been absorbed in the single thermal inertia parameter $I$, which is defined as

$$I = \sqrt{k \rho c_p} \tag{5}$$

It is worth noting that Eq. (5) is only valid at the surface and provided that thermal conductivity is constant, which is not strictly true (see below). However, thermal inertia is a convenient way to describe the reaction of surface temperatures to insolation changes, and it is thus widely used. In addition, lack of data usually does not allow different parameters to be disentangled, although recent studies indicate that the radiometrically measured thermal inertia changes with heliocentric distance for an individual object (Rozitis et al. 2017, Marsset et al. 2017). Nevertheless, temperature dependence of thermal conductivity can have a significant influence when interpreting conductivity in terms of regolith grain size (e.g., Gundlach and Blum 2013; Piqueux and Christensen 2011; Sakatani et al. 2017), and thermal inertia is therefore generally interpreted at a representative surface temperature (e.g., Müller et al. 2017; Gundlach and Blum 2013). In addition, care must be taken when converting thermal inertia to material parameters like thermal conductivity, since different combinations of material parameters govern the temperature at the surface (thermal inertia) and in the subsurface (thermal diffusivity).

There is evidence for regolith layering in terms of thermal conductivity from remote sensing data of main belt asteroids (MBA), and Harris and Drube (2016) argue that thermal inertia typically increases significantly as a function of thermal skin depth for these bodies. This effect may be caused by the fact that for slowly rotating asteroids surface temperature and thus surface thermal inertia is influenced by deeper layers than for fast rotators, and thus for slow rotators they probe regolith material parameters to greater depth. The observed increase of surface thermal inertia as a function of skin depth can thus be interpreted as thermal conductivity increasing as a function of depth, and it should therefore be kept in mind that a similar layering may be present on Ryugu.

## 3.1. Measured Quantities

### 3.1.1. Thermal Inertia and Albedo

Relevant thermal observations of Ryugu are summarized in Müller et al. (2017) and include: SUBARU-COMICS, multiple N-band (Aug 2007), AKARI IRC 15/24 micron photometry (May 2007), Herschel-PACS, 70/160 micron photometry (Apr 2013), Spitzer-IRS spectrum (5-38 micron) (May 2008), Spitzer-IRAC 3.6/4.5 micron two-epoch thermal light curves (Feb/May 2013), Spitzer-IRAC 3.6/4.5 micron multiple-epoch point-and-shoot sequence (Jan-May 2013), new data from NEOWISE at 4.6 micron (Sep 2016; J. Masiero, priv. comm.)

A thermal inertia of 150 to 300 J $m^{-2}$ $s^{-1/2}$ $K^{-1}$ is required to explain the available rich data set of thermal measurements over a wide phase-angle and wavelength range. The corresponding maximum surface temperatures are in the range ~320 to 375 K ($r_{helio}$ = 1.00 - 1.41 AU for our observational thermal data set), with $T_{max}$ ~350 K at the object's semi-major axis distance ($a$ = 1.18 AU). The corresponding shape model is very spherical with the volume-equivalent diameter 850-880 m (compare Sec. 2.1.2). The surface has thermal inertia of 150-300 J $m^{-2}$ $s^{-1/2}$ $K^{-1}$ and the roughness is small with the rms of surface slopes < 0.1, which means the surface appears very smooth in the radiometric context (connected to a very low width-to-depth ratio in the modelled spherical crater segments on Ryugu's surface). The reference model of Müller et al. (2017) has the pole direction (340°, - 40°), the volume-equivalent diameter 865 m, surface thermal inertia 200 J $m^{-2}$ $s^{-1/2}$ $K^{-1}$, low surface roughness with rms of surface slopes 0.05, geometric V-band albedo 0.049, infrared emissivity 0.9, and rotation period 7.63109 h (compare Sec. 2.1.1 and 2.1.2). Geometric albedo, phase integral, as well as the spherical bond albedo have already been discussed in Sec. 2.1.3, and values presented there are compatible with estimates from thermophyiscal models, which yield geometric V-band albedos of approximately 0.042 to 0.055. Using a phase integral of $q = 0.29 + 0.684G$ the bond albedo can be calculated using $A = qp_V$. The uncertainty in $G$ translates into an uncertainty in the phase integral $q$ (Bowell et al. 1989), and combined with a 5% accuracy of the $q - G$ relation (Muinonen et al. 2010) a bond albedo of A=0.019±0.003 for the full admissible thermal inertia and roughness range is obtained.

## 3.2. Derived Quantities

### 3.2.2. Thermal Conductivity

Thermal conductivity of particulate media is known to be very low in a vacuum, because the weak contacts between particles act as high thermal resistance points for heat conduction, and contributions of heat transfer by gas are absent (e.g. Fountain and West 1970). In fact, thermal conductivity of the lunar regolith was estimated to be less than 0.03 W m$^{-1}$ K$^{-1}$ even at depths below 50 cm based on the analyses of Apollo Heat Flow Experiment data (Langseth et al. 1976; Grott et al. 2010), and much lower values are expected at shallower depth. Accordingly, thermal conductivity of Ryugu's surface is also expected to be very low if it is covered by regolith or small particles.

Thermal conductivity of the top surface layer can be constrained using Eq. (5) if thermal inertia is assumed to be known from, e.g., ground-based observations. For Ryugu, using the estimated typical value of 200 J m$^{-2}$ K$^{-1}$ s$^{-1/2}$ (Müller et al. 2017) and assuming a bulk density of $\rho$ = 1100–1500 kg m$^{-3}$ (typically 1270 kg m$^{-3}$) as well as a specific heat capacity $c_p$ of 758 J kg$^{-1}$ K$^{-1}$ at 300 K (as an averaged temperature of daytime on Ryugu), thermal conductivity is constrained to be $0.020 < k < 0.108$ W m$^{-1}$ K$^{-1}$, with a most likely value of $k$ = 0.042 W m$^{-1}$ K$^{-1}$. However, it is worth pointing out again that this estimate is only valid if density, specific heat capacity, and thermal conductivity are constant as a function depth and time, which will in general not hold. Rather, on the actual Ryugu, these parameters may have depth-directional distributions and will vary with time due to the diurnal temperature variations.

Thermal conductivity of regolith-like powders is a complicated function of parameters such as, e.g., grain size, porosity, temperature, and particle surface energy, and different models have been proposed for this problem (e.g. Halajian and Reichmann, 1969; Hütter et al. 2008, Gundlach and Blum, 2012). Here we use the integrated model by Sakatani et al. (2017) that was developed based on experimental studies. Under vacuum conditions, effective thermal conductivity is given by the sum of "solid conductivity" originating from thermal conduction through inter-particle contacts and "radiative conductivity" originating from thermal radiation through void spaces between the particles, and it is given by

$$k = k_{\mathrm{solid}} + k_{\mathrm{rad}} \,. \tag{6}$$

The solid part of the conductivity $k_{\mathrm{solid}}$ can be modelled as serial and parallel connections of heat paths around the contact areas between spheres and it may be expressed as

$$k_{\mathrm{solid}} = \frac{4}{\pi^2} k_{\mathrm{m}} (1-\phi) C \xi \frac{r_{\mathrm{c}}}{R_{\mathrm{p}}} \,, \tag{7}$$

where $k_{\mathrm{m}}$ is the thermal conductivity of the bulk material, $\phi$ is the porosity, $C$ is the coordination number, $r_{\mathrm{c}}$ is the radius of the contacts area between two spheres, $R_{\mathrm{p}}$ is the radius of the grains, and $\xi$ is a coefficient that represents reduction of contact heat conductance due to surface roughness on the grains; $\xi = 1$ if particles are perfectly smooth spheres, and comparison with experimental data showed that $\xi \leq$ 1 for natural samples including spherical glass beads. The coordination number may be expressed as a function of porosity, and according to Suzuki et al. (1980) it is given by

$$C = \frac{2.812(1-\phi)^{-1/3}}{f^2(1+f^2)}, \tag{8}$$

where $f = 0.07318 + 2.193\phi - 3.357\phi^2 + 3.194\phi^3$. The contact radius $r_\mathrm{c}$ between two spheres depends on the force with which the particles are pressed together as well as their elastic parameters, and under the micro-gravity environment of Ryugu surface energy $\gamma$ and thus van der Waals forces will dominate the solid conductivity contribution. Contact radius $r_\mathrm{c}$ may be expressed as

$$r_\mathrm{c} = \left[\frac{3(1-\nu^2)}{4E}\left\{F + \frac{3}{2}\pi\gamma R_\mathrm{p} + \sqrt{3\pi\gamma R_\mathrm{p} F + \left(\frac{3}{2}\pi\gamma R_\mathrm{p}\right)^2}\right\}\right]^{1/3}, \tag{9}$$

where $\nu$ is Poisson's ratio, $E$ is Young's modulus, $F$ is the external compressive force on the particle, and $\gamma$ is the surface energy (Johnson et al. 1971). $F$ is given as the compressional stress $\sigma$ divided by the number of grains per unit cross-sectional area, and

$$F = \frac{2\pi R_\mathrm{p}^2}{\sqrt{6}(1-\phi)}\sigma. \tag{10}$$

On planetary surfaces, the compressional stress is caused by the self-weight according to $\sigma = \rho_\mathrm{m}(1-\phi)gz$ with $\rho_\mathrm{m}$ being the bulk density of the grain, $g$ gravitational acceleration, and $z$ depth, but on Ryugu this contribution will be small. Overburden pressure on Ryugu is only of the order of 0.01 Pa at a depth of 10 cm, and van der Waals forces will thus dominate. However, in total the contribution of the solid conductivity to the total conductivity will be small for typical NEAs with thermal inertia *I* larger than 100 J $m^{-2}$ $K^{-1}$ $s^{-1/2}$ when compared to the radiative contribution, and radiative conductivity will dominate.

The radiative part of the conductivity may be modeled by approximating the particle layers as multiple infinitely-thin parallel slabs. Then, radiative conductivity may be written as

$$k_\mathrm{rad} = \frac{4\epsilon}{2-\epsilon}\sigma_\mathrm{SB} L T^3, \tag{11}$$

where $\epsilon$ is the grain surface's thermal emissivity, $\sigma_\mathrm{SB}$ is the Stefan-Boltzmann constant, $T$ is temperature, and $L$ is the effective radiative distance between the slabs. Thus, $L$ is proportional to the geometrical clearance or typical pore length between the grains, and in a homogeneous packing of equal-sized spheres the geometrical pore length can be expressed in terms of grain size and porosity (Piqueux and Christensen 2009). Additionally introducing a factor $\zeta$ to scale the geometrical length to the effective one, $L$ is given by

$$L = 2\,\zeta\left(\frac{\phi}{1-\phi}\right)^{1/3} R_\mathrm{p}. \tag{12}$$

Comparison with experimental data for glass beads showed that $\zeta$ increases with decreasing grain size, which may be caused by a number of effects: First, the assumption of independent scatterers may break down, or a change of the scattering characteristics may take place for particles smaller than the thermal wavelength, resulting in more forward scattering. Further, radiative transfer may be enhanced across gaps smaller than the thermal wavelength (Rousseau et al. 2009), or there may be bias in the experimentally derived $k_{solid}$. The latter may be due to the fact that the temperature dependence of $k_{rad}$ is not exactly $T^3$ due to the coupling between solid and radiative heat transfer (Singh and Kaviany 1994), thus an extrapolation of measured data to 0 K for determining $k_{solid}$ may not be perfect.

A potential shortcoming of the above theory is that natural regolith does not consist of monodispersed spherical particles. However, laboratory measurements of the thermal conductivity of lunar regolith analogue material (Sakatani et al. 2018) show that measured thermal conductivity is in close agreement with theory if the median particle size (in the volume fraction sense) is used in the above equations. This is due to the fact that (a) the solid conduction only weakly depends on particle size and (b) the average pore size is close to the one calculated with the median radius using Eq. (12).

Another simplifying assumption made above is the spherical particle shape, while natural regolith is generally both non-spherical and angular. The deviation from the ideal shape can be measured by quantities called sphericity and angularity, and both shape indicators may influence the contact radii in the solid conduction contribution in complicated ways. However, their main contribution on thermal conductivity is to change regolith porosity, which is sufficiently described by the median particle radius. Thus, the current best estimate for determining $\zeta$ is an empirical relation derived from the experimental results of Sakatani et al. (2018) for a lunar regolith simulant, and it is given by

$$\zeta = 0.68 + \frac{7.6 \times 10^{-5}}{R_\mathrm{p}}. \tag{13}$$

In the above model, thermal conductivity of the regolith depends mainly on particle size and porosity, and ranges for these two parameters can be estimated if thermal inertia or effective thermal conductivity is known. Figure 6 shows a thermal inertia contour plot as a function of particle size and porosity. For thermal inertia values of $I$ = 150–300 J $m^{-2}$ $K^{-1}$ $s^{-1/2}$ as derived from remote sensing observations (Müller et al. 2017) a typical particle size for the Ryugu regolith can be estimated to be 3–30 mm, with 6–10 mm being the most likely. In this range of thermal inertia, the radiative conductivity is dominant over the solid conductivity, so that uncertainty of assumed parameters in the solid conductivity model, such as Young's modulus and surface energy, does not affect the grain size estimation. On the other hand, porosity cannot be effectively constrained because a major part of heat flow is made by the radiative heat transfer in this particle size range. Radiative thermal conductivity increases while the heat capacity decreases with porosity, therefore this balance of both parameters makes thermal inertia nearly constant for various porosities.

For a verification of the determined grain size, a comparison with the thermal model developed by Gundlach and Blum (2013) can be made. This model is based on the same physical concept as the calculations described above, however deviates in some details. Basically, these differences are: (i) the formulation of the network heat conductivity, (ii) the description of the radiative thermal transport process, and (iii) the irregularity parameter measured by different calibration measurements. A derivation of the regolith grain radius based on the Gundlach and Blum (2013) formulations yields radii ranging between 1.1 and 2.5 mm (for volume filling factors between 0.1 and 0.6). Thus, both models agree rather well, with the Gundlach and Blum (2013) model having slightly smaller sizes, which provides confidence in the grain-size determination from thermal inertia measurements of planetary surfaces.

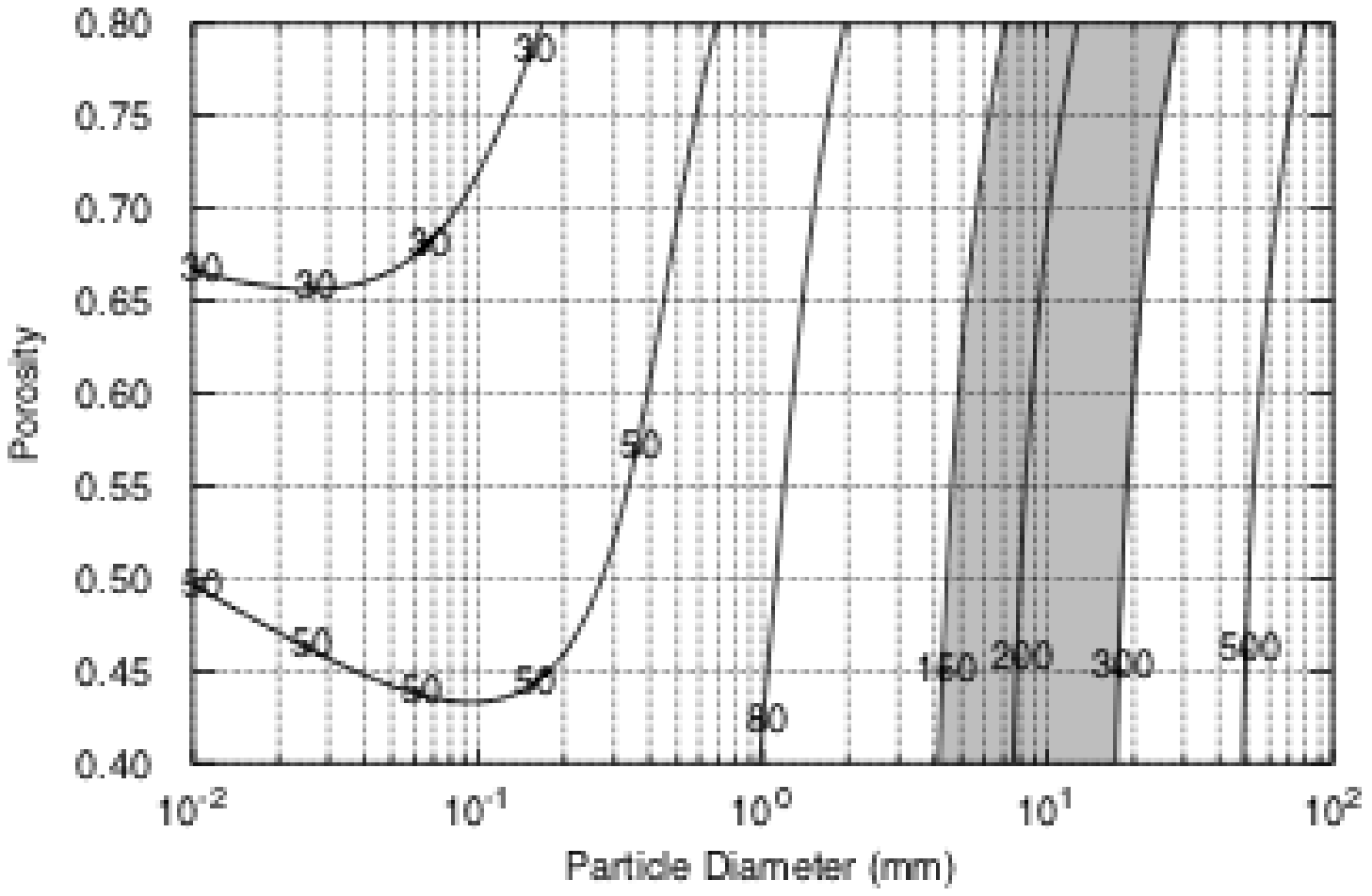

**Figure 6: Contour plot of thermal inertia (in units of J m$^{-2}$ s$^{-1/2}$ K$^{-1}$) as a function of particle diameter and porosity. The gray-filled region indicates the thermal inertia range estimated for Ryugu by ground based observations (Müller et al. 2017). The regolith parameters assumed in the calculation are $c_{\mathrm{p}}$ = 758 J kg$^{-1}$ K$^{-1}$, $T$ = 300 K, $\rho_{\mathrm{m}}$ = 3110 kg m$^{-3}$, $k_{\mathrm{m}}$ = 0.0021 $T$ + 1.19 W m$^{-1}$ K$^{-1}$, $\gamma$ = 0.032 J m$^{-2}$, $\nu$ = 0.25, $E$ = 78 GPa, $\epsilon$ = 1.0, $\xi$ = 0.12, $z$ = 0.01 m, and $g$ = 1.8 x 10-4 m s$^{-2}$, respectively. $\gamma$ and $\xi$ were optimized for experimental data of lunar regolith simulant JSC-1A (Sakatani et al. 2018)**

## 3.3. Predicted Properties

### 3.3.3. Regolith Heat Capacity

The specific heat capacity of rocks and soils at low temperatures has been studied for lunar samples (Robie et al. 1970; Fujii and Osako 1973; Hemingway et al. 1973), and a strong temperature dependence has been found. The suite of materials studied includes particulate material such as lunar fines and soils, but brecciated lunar rocks as well as basalts have also been studied. A best fit to the lunar soils data is given by Hemingway et al. (1973) and the specific heat capacity can be approximated as

$$c_P = -23.173 + 2.127T + 1.5009 \cdot 10^{-2}T^2 - 7.369910^{-5}T^3 + 9.6552 \cdot 10^{-8}T^4, \qquad (14)$$

where $c_p$ is specific heat capacity in units of J kg$^{-1}$ K$^{-1}$ and $T$ is temperature in K. This best fitting formula is accurate to within 2 percent down to temperatures of 200 K and to within 6% down to temperatures of 90 K. At higher temperatures, the expression

$$c_P = 4184 \cdot (0.2029 \; + \; 0.0383 \left( 1 - \exp\left(\frac{350 - T}{100}\right)\right), \tag{15}$$

by Wechsler et al. (1972) may be used (also see Ledlow et al. (1992), Stebbins et al. (1984), and Schreiner et al. (2016)). The fit to the data is shown along with the data in Figure 7 and extends to temperatures of up to 350 K, as appropriate for Ryugu. The extrapolation applicable at higher temperatures is shown as a dashed line.

Measurements on lunar material are in good agreement with measurements performed by Winter and Saari (1969) using other geological materials as well as measurements of the specific heat capacity of meteorites performed by Yomogida and Matsui (1983) as well as Consolmagno et al. (2013). It may be worth noting that a trend exists with respect to the iron content of the considered samples, with low iron corresponding to high heat capacity (Yomogida and Matsui 1983). While heat capacity thus shows a strong temperature dependence, this is only relevant if the near surface regolith layer is considered. At depths below a few skin depths, perturbations rapidly decay such that the regolith can be assumed isothermal for the purpose of determining its heat capacity. Evaluated at a representative average daytime surface temperature of 300 K, a value of $c_p$ = 758 J kg$^{-1}$ K$^{-1}$ is obtained by evaluating Eq. 14.

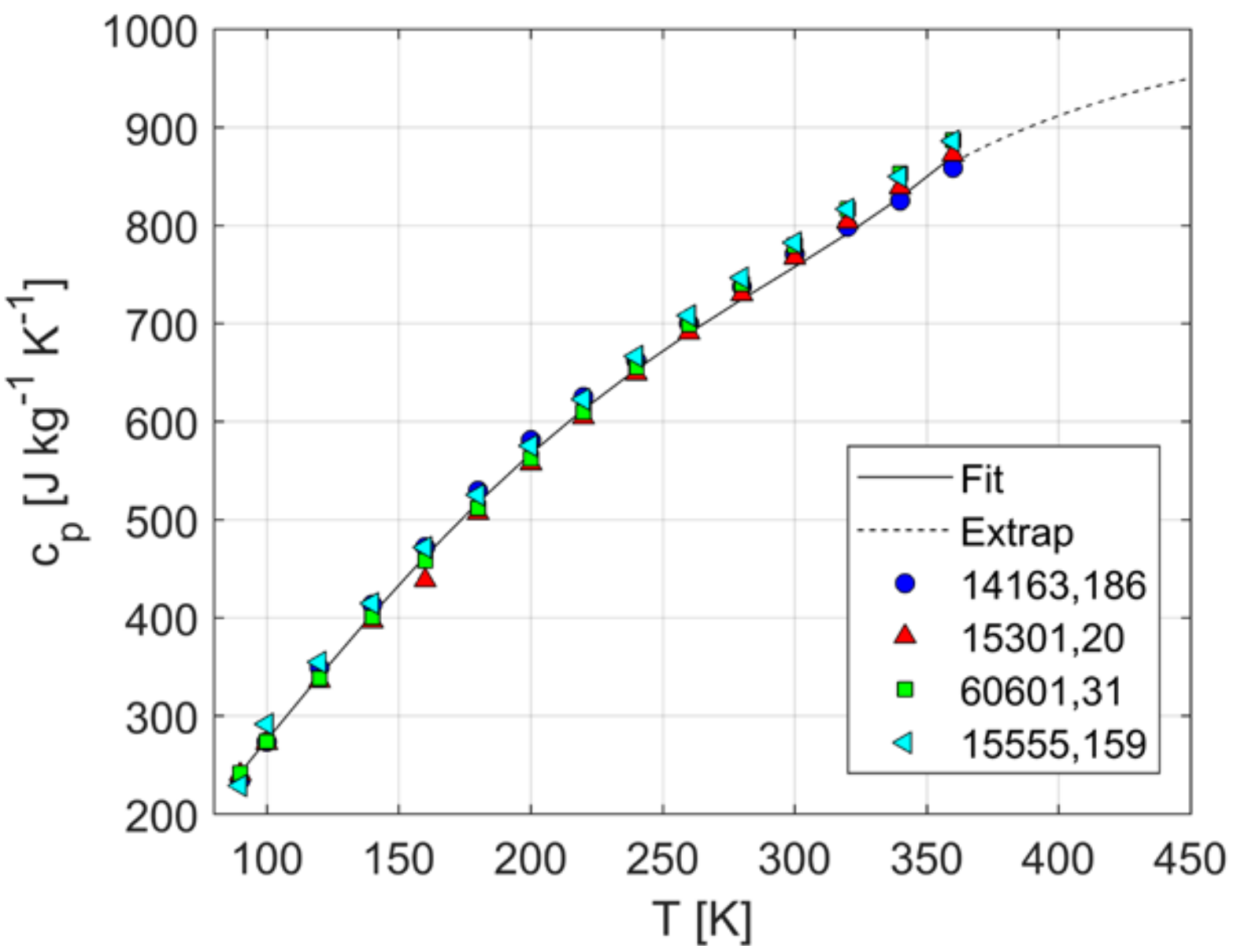

**Figure 7: Specific heat capacity of lunar samples 14163,186 (fines >1 mm, blue circles), 15301,20 (soil, red triangles), 60601,31 (soil, green squares), and 15555,159 (basalt, cyan left triangles) as a function of temperature together with the best fitting curve (solid line) as well as an extrapolation for temperatures beyond 350 K (dashed line). Data and fit from Hemingway et al. (1973), extrapolation from Ledlow et al. (1992).**

### 3.3.4. Surface Temperatures, Thermal Gradients, and Thermal Fatigue

The thermal evolution of Ryugu is simulated by solving Eq. (3) numerically. In general the input flux at the surface is not only the heat flux from the sun but also the sum of the heat radiation from the other surface area (Rozitis and Greeen 2011; Davidsson and Rickman 2014; Davidsson et al. 2015). However, Ryugu is predicted to be a round-shape body from ground-based observations. Thus hereafter, for simplicity, it is assumed that the surface is convex and areas to be calculated do not receive heat radiation from other areas.

It is to be noted that the thermal emissivity is not needed to be the complement of the surface albedo, because the peak wavelength of the solar radiation is in the range of the visible light while that of the thermal radiation from the surface of an asteroid is around or longer than 10 um. Thus the thermal emissivity and the albedo are independent parameters. We adopt 0.019 for the albedo and 0.90 for the emissivity of the surface of Ryugu as nominal values. Hereafter we will show the equilibrium temperature evolution as a function of local time, which is obtained after 100 asteroid rotations at fixed solar distance and sub-solar latitude. The difference between the equilibrium temperature distribution and a full numerical simulation taking into account the orbit of the body around the Sun is less than a few K.

Figure 8 shows the evolution of equilibrium surface temperature as a function of local time with constant thermal inertia (left panel, thermal inertia equal to 200 J m$^{-2}$ s$^{-1/2}$ K$^{-1}$) but various local latitudes, and constant local latitude (right panel, 0 degree) but various thermal inertias. In these numerical simulations we adopt a solar distance of 1.2 AU and the sub-solar latitude of 0 degree (equator) for simplicity. As shown in Figure 8, surface temperature changes with solar direction and decreases almost linearly during night time. Both the maximum and minimum temperatures decrease with latitude for a constant thermal inertia. The local time of the maximum temperature is achieved after noon and the degree of phase shift relative to the noon is almost the same for each latitude. On the other hand, as shown in the right panel of Figure 8, the local time of the maximum temperature becomes later with thermal inertia. The phase shift is a diagnostic parameter to estimate the thermal inertia from the surface temperature distribution obtained by the instrument TIR (Takita et al. 2017). The maximum temperature decreases with thermal inertia and the minimum temperature increases with thermal inertia because subsurface temperatures are kept warmer for larger thermal conductivities. It is to be noted that the total thermal radiation from the surface must balance with solar influx in the equilibrium state. Thus the decrease of maximum temperature is smaller than the increase of minimum temperature since the thermal radiation is proportional to the fourth power of the temperature.

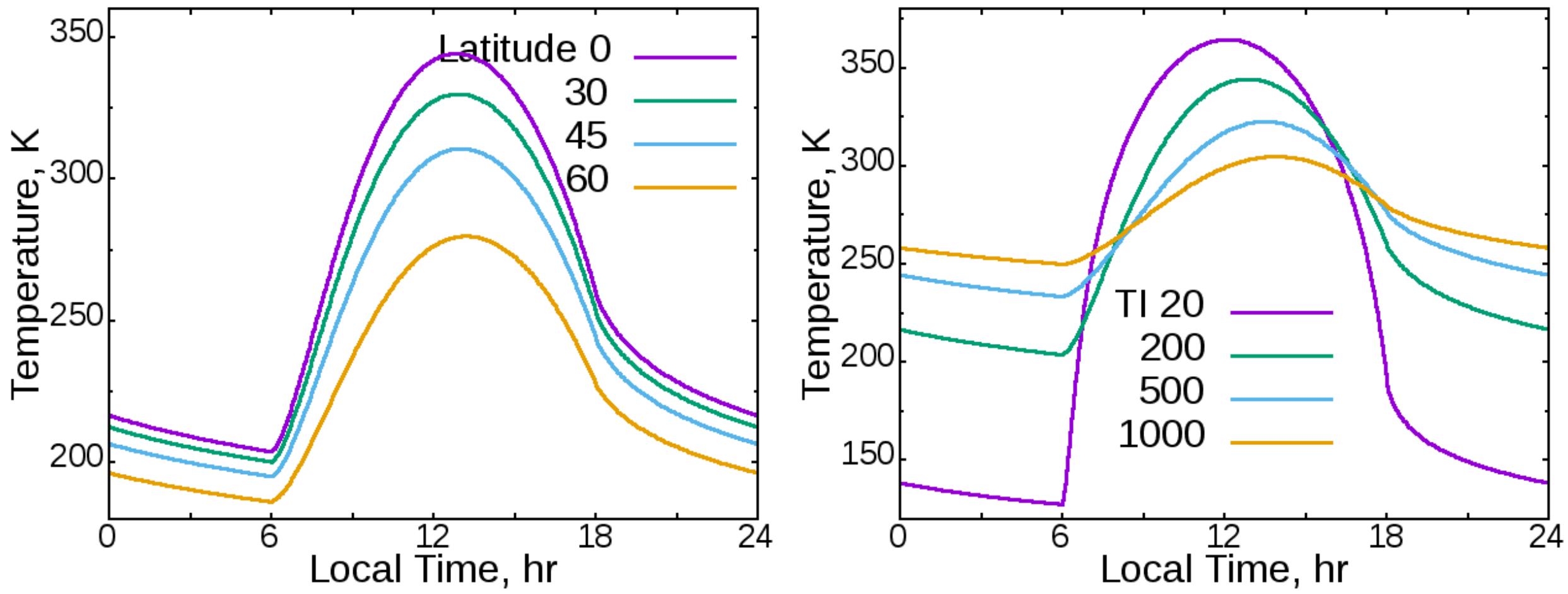

**Figure 8: Surface temperature change as a function of local time for various latitudes (left) and thermal inertia (right). The sub-solar latitude is assumed to be 0 (equator).**

Figure 9 shows results similar to Figure 8, but for a spin axis tilted by 30 degrees. In this case the duration of day and night depends on latitude. As a result the maximum temperature is not needed to be obtained at the latitude identical to the sub-solar latitude. For the case with a thermal inertia of 1000 J $m^{-2}$ $s^{-1/2}$ $K^{-1}$, the maximum temperature is obtained at the latitude of 45 degrees. Despite the change of the length of day time, the shift of the local time of peak temperature is independent of the latitude. These results are similar to those obtained by Müller et al. (2017), who calculated maximum surface temperatures in the range of 320 to 375 K by considering the derived object properties and heliocentric distances of 1.00 - 1.41 AU. At the object's semi-major axis distance of 1.18 AU, Müller et al. (2017) find a reference maximum temperature of 350 K.

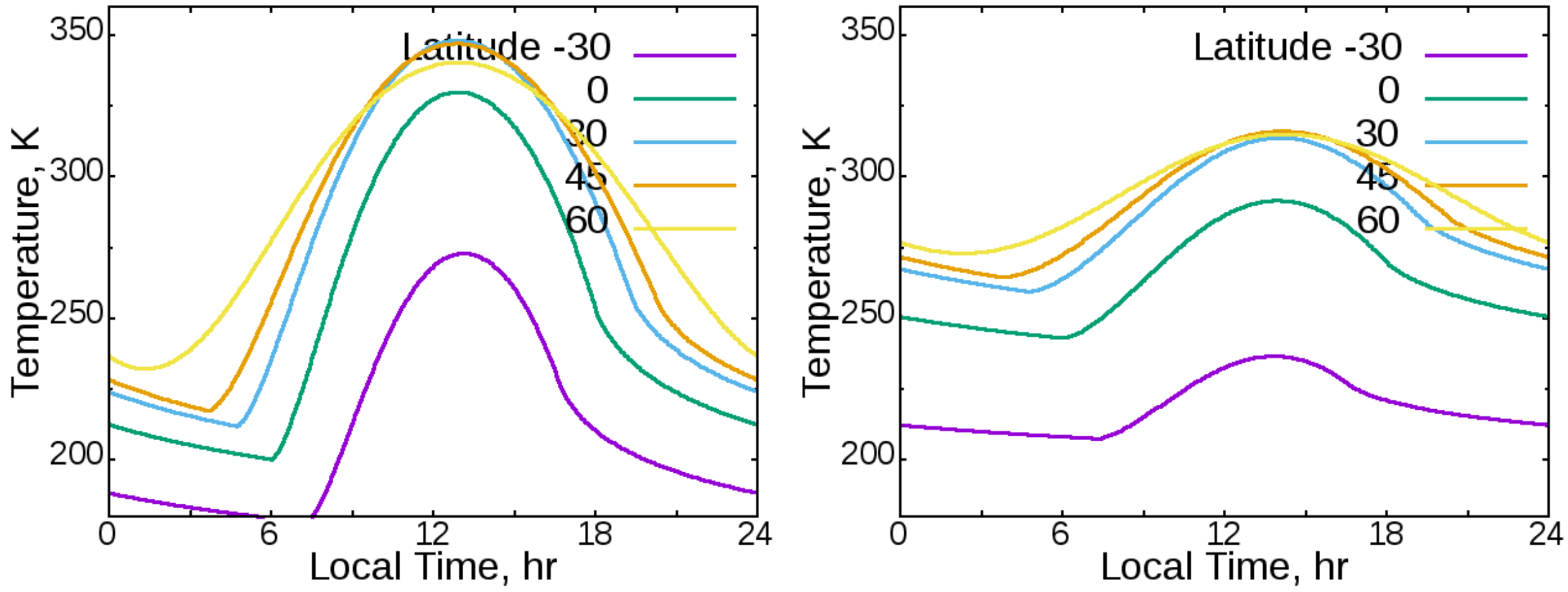

**Figure 9: Left: Similar to Figure 8, but for a case with sub-solar latitude of 30, i.e. a tilted spin axis. Left: The thermal inertia is 200 J $m^{-2}$ $s^{-1/2}$ $K^{-1}$. Right: Thermal inertia is 1000 J $m^{-2}$ $s^{-1/2}$ $K^{-1}$.**

Delbo et al. (2014) suggested that thermal fatigue could be the driver of regolith formation on asteroids with perihelion distances of less than 2.5 AU, including NEAs such as Ryugu. Thermal fatigue describes the breakdown of larger boulders due to thermal stress. The regolith is exposed to large temperature differences between night and day time temperatures causing large temperature gradients and therefore stress in the material leading to crack formation and propagation. Delbo et al. (2014) estimated that thermal fatigue erodes the surfaces of NEAs like Ryugu several orders of magnitude faster than erosion caused by micrometeoroids. Consequently, studying the temperature gradient at the surface might be essential for understanding the evolution of the surface of Ryugu. Figure 10 represents the evolution of the temperature gradient just beneath the surface as a function of the local time and latitude. The thermal inertia and sub-solar latitude are assumed to be 1000 J $m^{-2}$ $s^{-1/2}$ $K^{-1}$ and 0 degree, respectively. We adopted a larger thermal inertia than the nominal value of Ryugu because dust creation due to the thermal fatigue takes place at the surface of consolidated boulders or monolithic rocks. As shown in Figure 10 the maximum temperature gradient is achieved before noon. The absolute value of thermal gradient is larger for lower latitude. This is simply because the larger thermal spatial gradient is obtained at the time when the maximum temporal gradient of the surface temperature is achieved. In fact, as shown in the right panel of Figure 8, the maximum temporal gradient at the surface is achieved before noon. The maximum thermal spatial gradient becomes larger for smaller thermal inertia. On the other hand, the negative maximum temperature gradient is achieved at sunset. After sunset the temperature gradient becomes smaller because the underground temperature distribution is alleviated during night time. The right hand panel of Figure 10 is similar to the left one, but uses a spin axis tilted by 30 degrees. In this case the maximum thermal gradient is achieved not only at the sub-solar latitude but at the sub-solar latitude and the equator. Thus the determination of location at which the maximum temperature gradient is achieved is not straightforward (Hamm et al. submitted). It is to be noted that the integration of thermal gradient through a day should be zero in the equilibrium state if the thermal conductivity is independent of temperature.

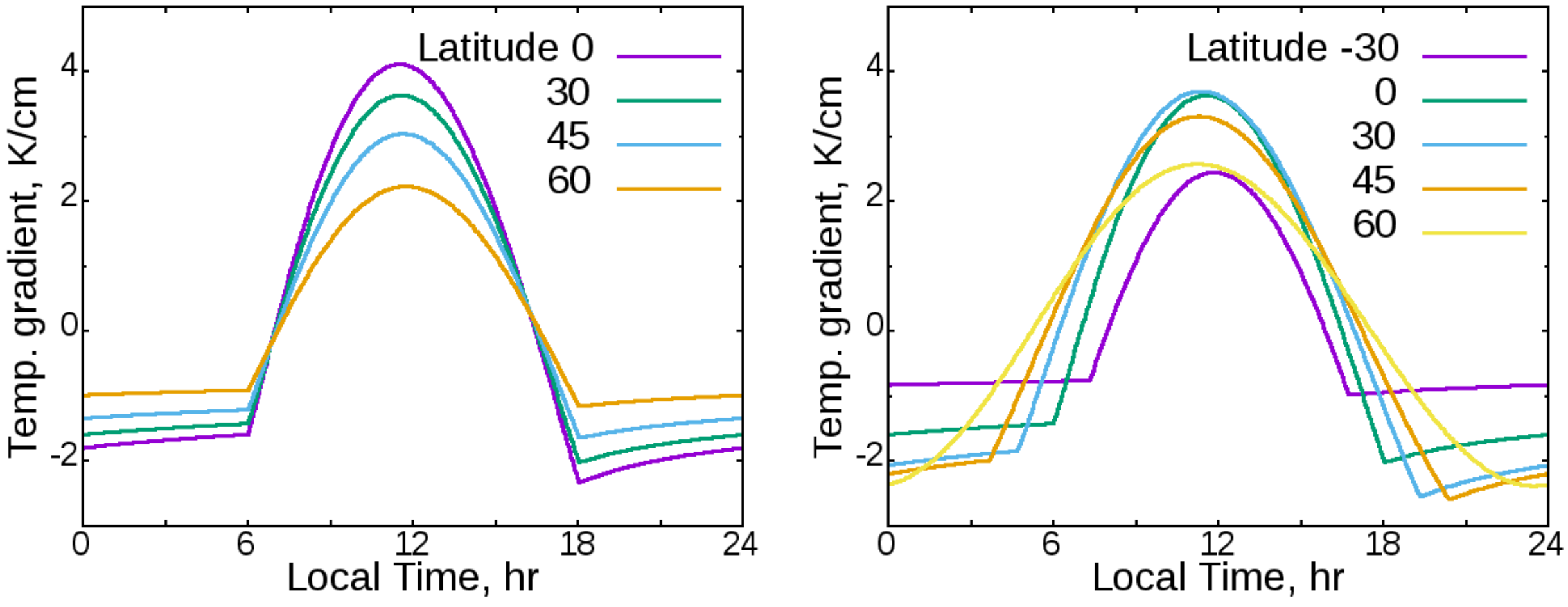

**Figure 10: Thermal gradient just beneath the surface as a function of local time. Left: Sub-solar latitude of 0 degree. Right: Sub-solar latitude of 30 degrees. The negative maximum of the thermal gradient is achieved at sunset while the positive maximum of the thermal gradient is achieved just before noon.**

We assumed above that the surface of Ryugu is smooth. If the roughness of the surface is taken into account, the apparent temperature distribution changes depending on the degree of roughness (Rozitis and Green 2011; Davidsson et al. 2014, 2015). The temperature of a slope is especially affected at dawn and dusk. A wall of a valley observed at a low phase angle will have a much higher temperature than the horizontal, surrounding surface. Contrarily, when observing a wall against the sun it will be much colder than its surroundings. In addition to the case of concave areas such as craters, the absorption of radiation could raise the temperature of the bottom area of the concavity depending on its size and shape. Moreover, we must point out that the bond albedo of a rough surface changes with solar incident angle (Senshu et al. in preparation). These effects need to be taken into account when analyzing the apparent temperature distribution obtained by the Hayabusa2 TIR instrument.

# 4. Regolith Mechanical Properties

## 4.1. Measured Quantities

There are no directly measured quantities of the mechanical properties of the regolith on Ryugu, as for any asteroid that has not been visited by a spacecraft. Therefore, the only quantities that can be determined must be derived from current data or rely on pure assumptions.

## 4.2. Derived Quantities

The only quantity that can be derived from observations is the dominant grain size, based on the measurement of the thermal inertia of the asteroid (see previous sections). The grain shape can also be derived from the only asteroid from which regolith particles have been returned, i.e. the asteroid Itokawa. However, we note that Itokawa belongs to the S taxonomic class, and its composition is thus very different from that expected for Ryugu. This probably also means that its mechanical properties are different, and therefore how regolith is generated on this body, as well as its properties, are likely to be different too. But since this is the only data point that we have so far, we use it in this section.

### 4.2.1. Grain Size

As summarized in Sec. 3, a thermal inertia of 150 to 300 J $m^{-2}$ $s^{-0.5}$ $K^{-1}$ is required to be compatible with remote sensing observations (Müller et al. 2017). Using the model by Sakatani et al. (2016), this translates to typical particle diameters of 3–30 mm for Ryugu, with 6–10 mm being the most likely. This is consistent with, although slightly larger, than the results by Gundlach and Blum (2013), who obtain grain diameters of 2.2 to 5 mm.

### 4.2.2. Grain Shape

For the grain shape as shown in **Figure 11** and **Figure 12**, we note that the mean aspect ratios of particles returned from the asteroid Itokawa by the Hayabusa mission, $b/a$ and $c/a$ ($a$ is the longest axis, $b$ is the medium axis, $c$ is the shortest axis) (Tsuchiyama et al. 2014), are $0.72 \pm 0.13$ and $0.44 \pm 0.15$, respectively, and are similar to the mean axial ratio of fragments generated in laboratory impact experiments ($a$:$b$:$c$ ≈ 2:√2:1 or $b/a \approx 0.71$ and $c/a \approx 0.5$). Note that the bulk density of Itokawa samples is 3.4 g $cm^{-3}$. The average porosity of Itokawa samples is 1.5%. We do not have any data yet for a C-type asteroid like Ryugu.

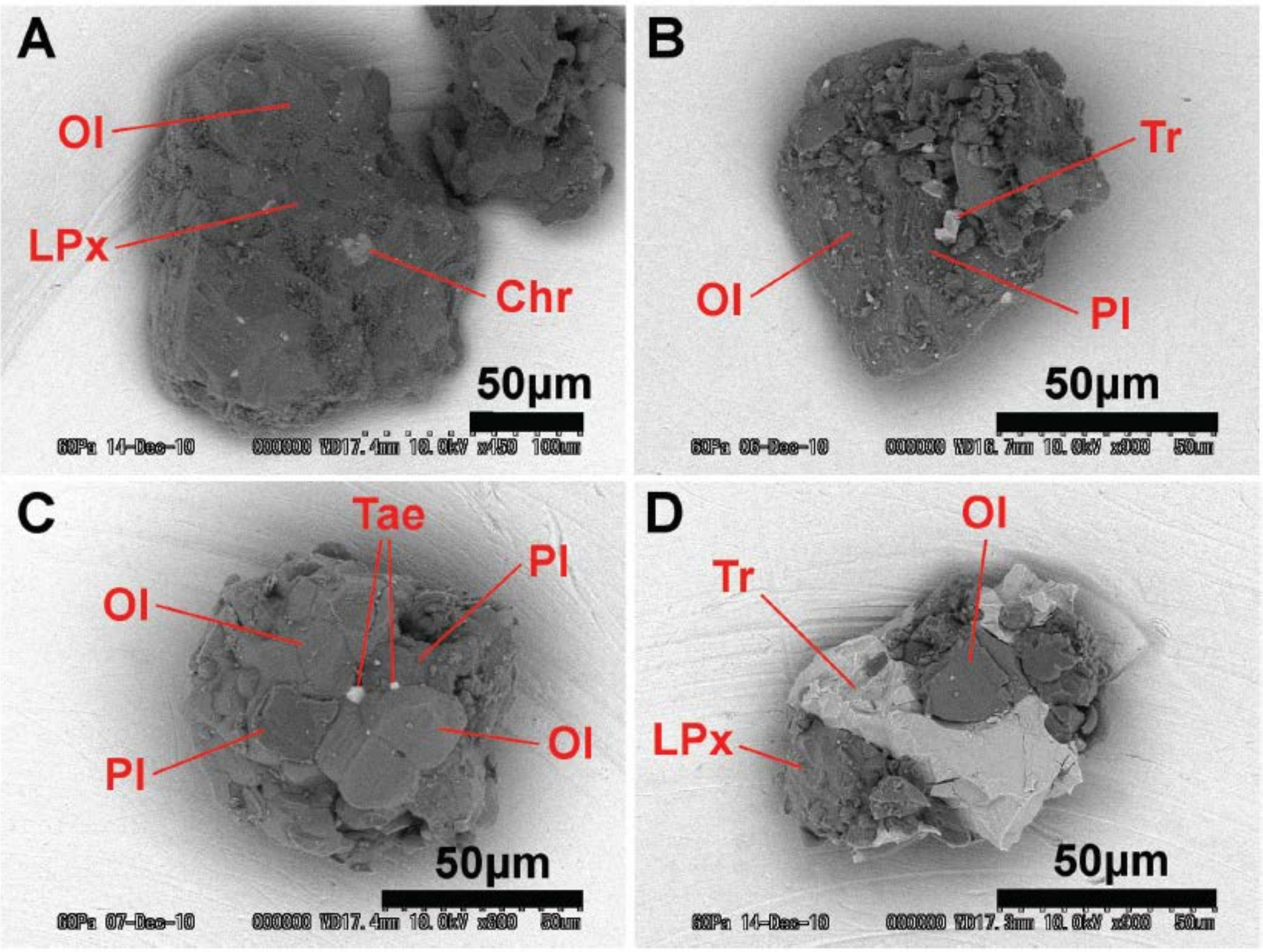

**Figure 11: Backscattered electron (BSE) images of asteroid Itokawa regolith grains RA-QD02-0030 (A), RA-QD02-0024 (B), RA-QD02-0013 (C), and RA-QD02-0027 (D) from (Nakamura et al. 2011). Here, Ol: olivine, LPx: low-Ca pyroxene, Chr: chromite, Pl: plagioclase, Tr: troilite, Tae: taenite.**

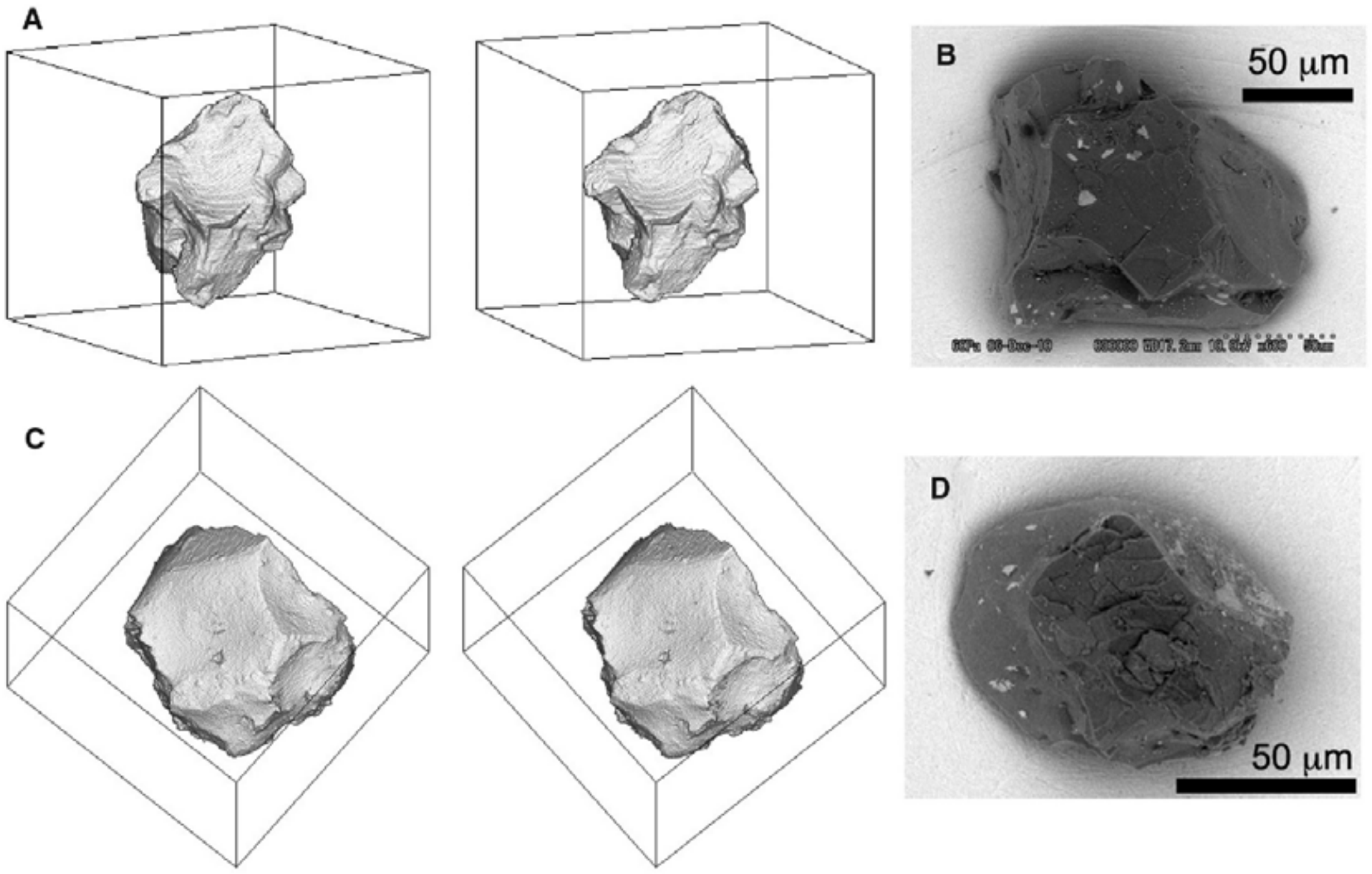

**Figure 12: The 3D external shapes of Itokawa particles. (A) Stereogram (box size 232 µm by 232 µm by 203 µm) and (B) SEM micrograph of RA-QD02-0023.(C) Stereogram (box size, 112 µm by 112 µm by 93 µm) and (D) SEM micrograph of RA-QD02-0042.**

## 4.3. Predicted Quantities

Because we do not yet have any detailed image of a C-type asteroid of Ryugu's size and any experiment done on such a surface allowing us to understand how it responds to an external action, most of the information regarding regolith mechanical properties of Ryugu are predictions.

### 4.3.3.Regolith Porosity

Natural regolith is typically non-spherical (measured by a quantity called sphericity) and angular (measured by "angularity"). Together with the adhesion properties, the median grain size and the size distribution determine the porosity.

It is well known that random packings of monosized coarse spheres have a porosity of 0.363 (dense) to 0.42 (loose packing). Adhesion increases the porosity, to the limit of 1; the decisive parameter is the granular Bond number $B_o = F_{vdw}/F_g$, i.e. the ratio of van der Waals forces and weight acting on a grain (Kiuchi and Nakamura 2014). If the particles are polydisperse, the porosity tends to decrease with the width of the size distribution (the smaller particles filling up the voids of the larger particles). Finally, non-spherical grains tend to increase the porosity compared to packings consisting of spherical grains.

A well-known packing theory which can be used for predictions is described in Zou et al. (2011). Here the equation expressing the "initial porosity" as a function of grain size has to be generalized to arbitrary gravity (instead of 1 g) and assumption has to be made on the increase of initial porosity due to shape effects.

Ryugu has a very low gravity so that compaction by lithostatic pressure is not likely to play a role in the upper meter or so. This is different from the Moon: the porosity of lunar regolith (as a function of depth) is relatively well known; it is 83% on the very surface (Hapke and Sato 2016), decreases quickly to about 58% at a few cm and to about 38% at a few meter depth. Apart from the very upper surface, we see compaction by overburden pressure at work (Schräpler et al. 2015; Omura and Nakamura 2017).

Given these considerations, a porosity of ~0.4-0.5 can be assumed for the near-surface regolith, which corresponds to the random loose packing of slightly adhesive [relative to weight!] cm-sized irregular but roundish particles with a size distribution.

#### 4.3.4. Regolith Subsurface Properties

According to Britt and Consolmagno (2001), materials in porous asteroids may be sorted by particle size. The large irregular pieces (and larger voids/fractures) may be located deeper inside the asteroid and the fine particle size fractions that are observed on their surfaces are restricted to the surface regolith zone (Figure 13). The large interior voids/fractures are preserved from infilling by the effects of friction on the smaller size fractions. Friction tends to dominate the downward pull of gravity and prevents the fine fractions from filtering into the interior of the asteroid and infilling the large fractures and voids. Friction may also play a role in allowing shattered asteroids to maintain their relief features and shape by resisting the movements of pieces within the object, in effect providing strength to non-coherent objects.

However, another mechanism, called the Brazil nut effect, may cause an opposite trend (Matsumura et al. 2014; Maurel et al. 2017) and segregate larger particles to the surface, as seen on Itokawa. Thus, coarse gravel (1 cm or greater) may be expected to exist on the surface overlaying fine grained material with the fraction of fines increasing with depth. The combination of low surface acceleration and solar radiation pressure tends to strip off fine particles that have been generated by comminution processes, and leave lags of larger, harder to move materials. On the other hand, this material may experience thermal fatigue that can turn them in smaller pieces. There is thus a competition between various processes, and we are not yet in a position to understand which one is dominant or their respective contributions to the final product. We would actually need at least one image of the surface of a C-type asteroid of Ryugu's size, which does not yet exist, and thus we have to wait for Hayabusa2 to provide one.

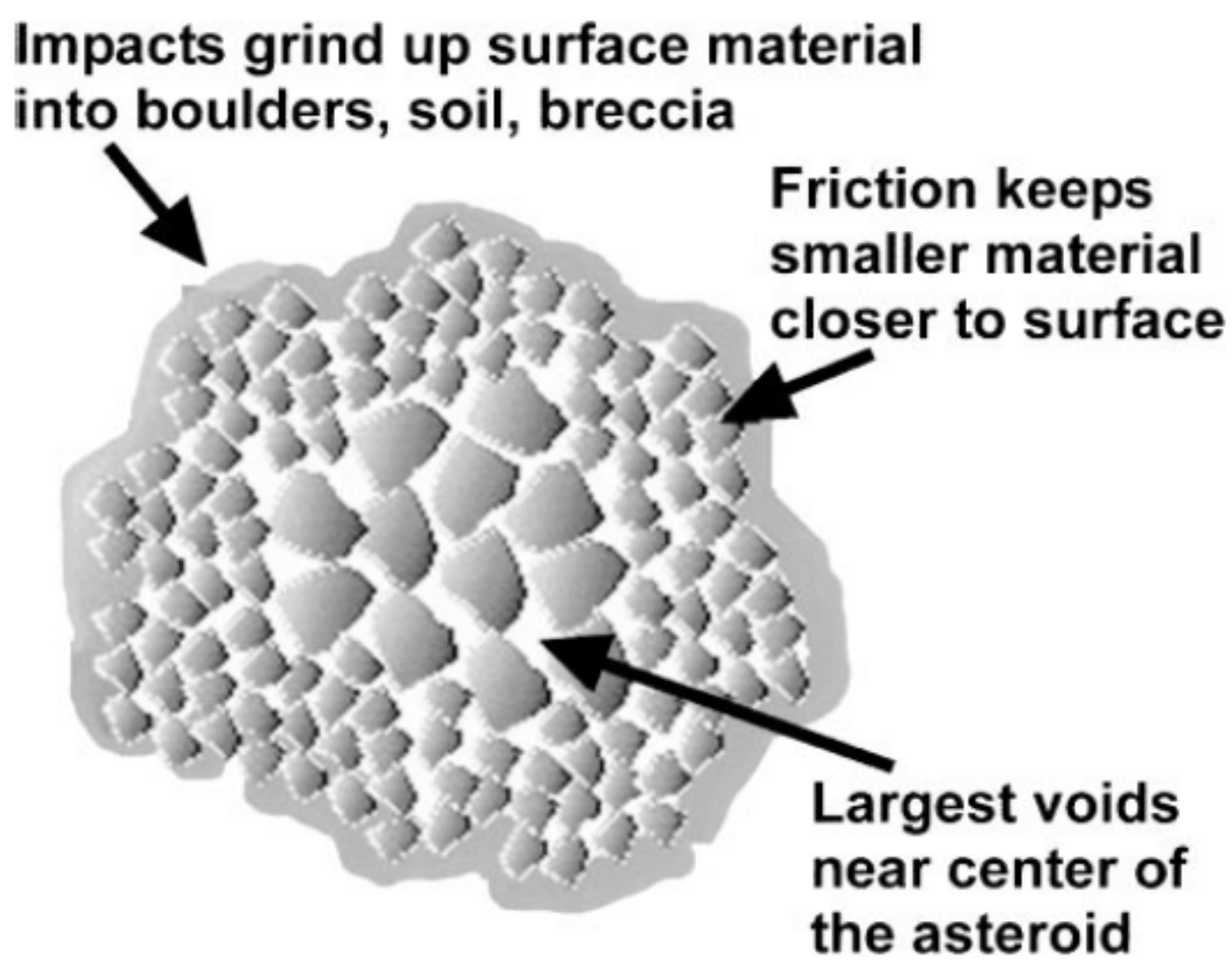

**Figure 13: Schematic of the possible structure of rubble pile asteroids. The original reaccretion of these objects is assumed to be particle size sorted with the largest pieces accreting first to form the core of the object and successively smaller sizes accreting later (Michel and Richardson, 2013). The finest size fractions are generated at the surface by regolith action. These fine particles are confined to the near-surface zone by the dominance of frictional forces (from Britt and Consolmagno 2001).**

#### 4.3.5. Regolith Cohesion, Angle of Friction, Tensile Strength and Their Influence on Shear Strength

In terrestrial environments, the shear strength of regolith is driven by inter-grain frictional forces and rotations, however both effects are largely affected by the level of gravity. Thus as gravitational forces decrease, other forces start to govern the behaviour of granular matter. With the decrease of gravitational forces, the importance of attractive forces like van der Waals forces rise. These attractive forces may, like higher gravity, increase the frictional forces by holding grains together and hence increasing the normal pressure. Thereby both frictional forces and resistance torques are increased, which in turn increase shear strength. Taking these considerations into account, the behaviour of granular matter gets even more complex in micro-gravity (μg). Furthermore, each particle is affected by the spin of the asteroid (particularly if it is fast) and the resulting forces, which constantly try to separate the grains and thus break the soil.

Cohesion, friction angles and relative density have been linked for the lunar regolith as shown in Figure 14.

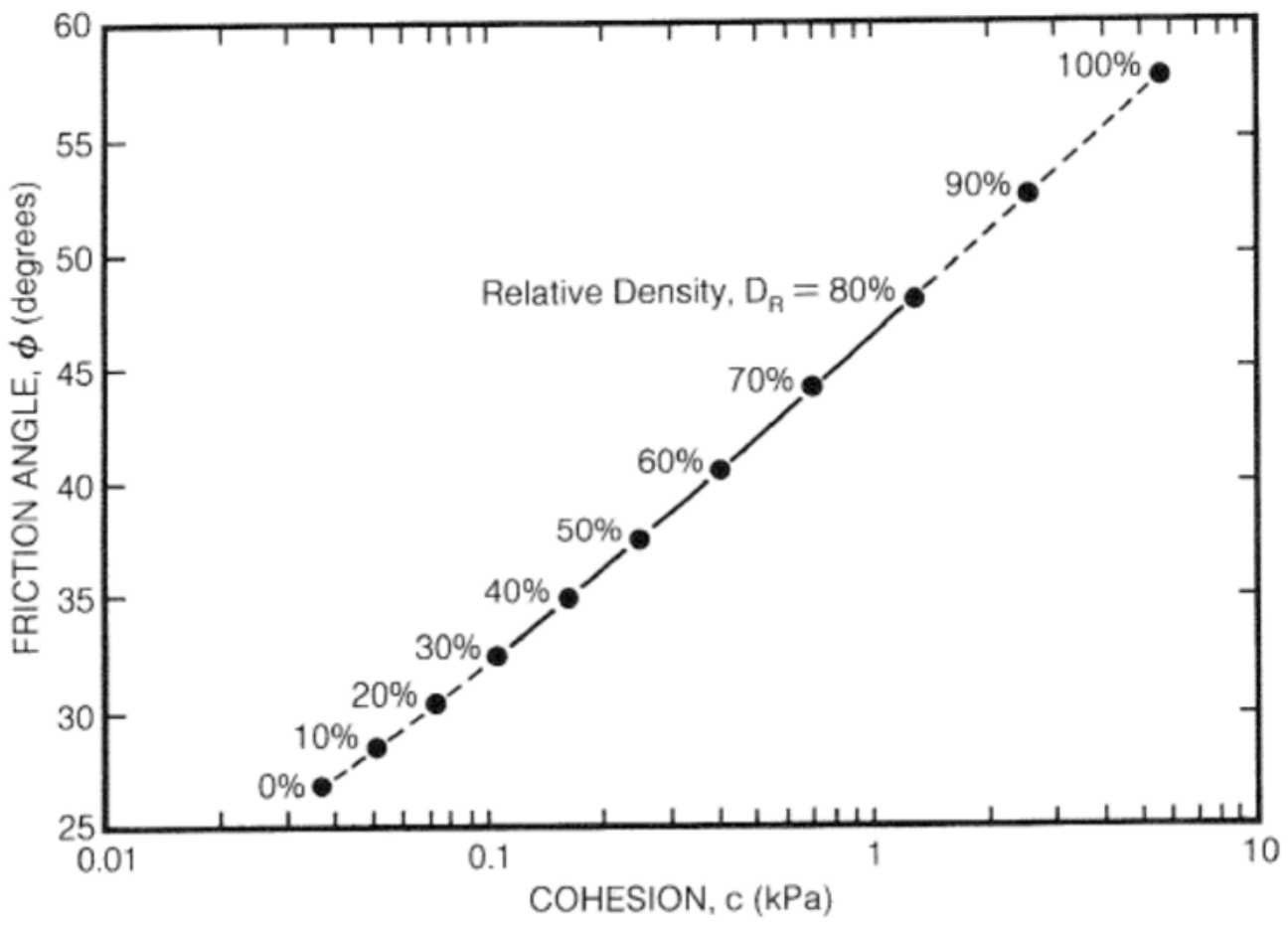

**Figure 14: Friction angle as a function of cohesion and relative density for the regolith on the Moon (same as Fig. 9.27 of Heiken et al. 1991).**

With an assumption of 50% porosity, the cohesion should be 300 Pa, and the angle of friction about 37 degrees. However, this is estimated for lunar regolith with fine grains. If Ryugu's regolith is coarser (1 mm-1 cm grains), then it may have lower cohesion, say 100 Pa (30% relative density) equivalent to an angle of friction of 33 degrees. From this, we can estimate the tensile strength from the Mohr-Coulomb curve to be 1 kPa or less (see Figure 9.26 of Heiken et al. 1991). The effect of gravity is a complicating factor, but Kleinhans et al. (2011) have shown that if gravity decreases the static angle of repose of regolith increases.

While cohesion in regolith is best known for the influence of water bridges in terrestrial environments, here the term cohesion covers the whole range of attractive forces. Due to the absence of water the most important attractive forces for μg environments are van der Waals forces, electrostatic forces, magnetostatic forces, and self-gravity (Scheeres et al. 2010; Murdoch et al. 2015). In a study comparing the cohesive forces to gravitational forces in different levels of gravity, Scheeres et al. (2010) have shown that cohesion might be one of the governing effects for the shear strength of granular matter in micro-gravity. Using scaling laws, the study showed that cohesion mostly governed by van der Waals forces may be able to be the governing force even for gravel-sized particles (~centimeter size) in Ryugu's ambient gravity as shown in Figure 15.

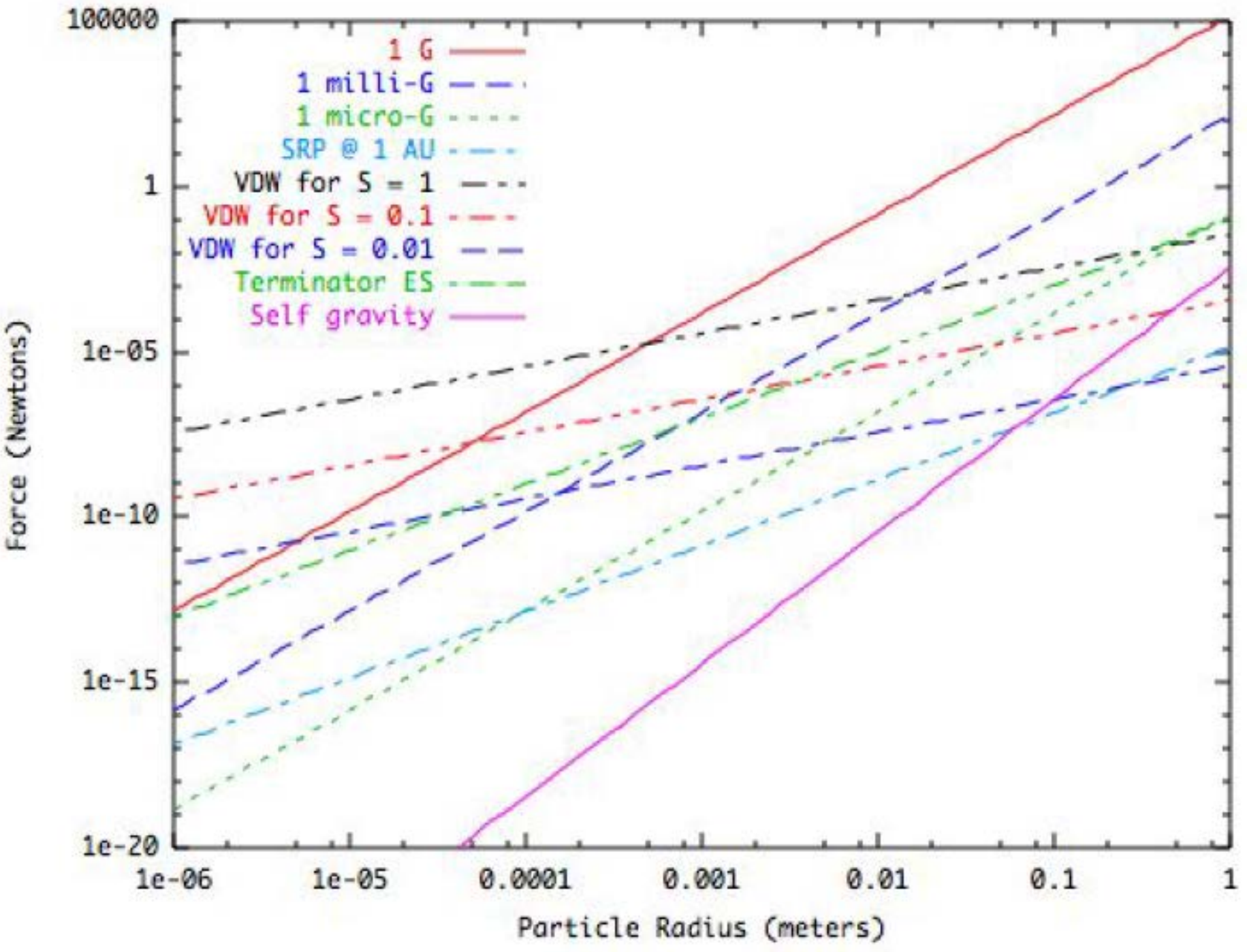

**Figure 15: Magnitude of different forces compared to gravity as a function of particle radius (from Scheeres et al. 2010). VDW expresses van der Waals force and S is the surface cleanness (S goes to 1 for clean surfaces). Ryugu's ambient gravity is estimated of the order of 0.25 milli-G.**

The non-dimensional bond number can be expressed by the relation between the gravitational weight and the cohesional forces (Scheeres et al. 2010, see Figure 16):

$$B = \frac{F_C}{F_g} \; . \qquad (16)$$

For the expected gravitational level of Ryugu, Scheeres et al. (2010) predict bond numbers of one for gain sizes up to $10^{th}$ of cm. They also state that clumping and macroscopic cohesional behaviour will be present even in regolith made of coarse sand portions. Other authors like Durda et al. (2012) studied the likely behaviour of asteroid regolith using terrestrial laboratory simulants.

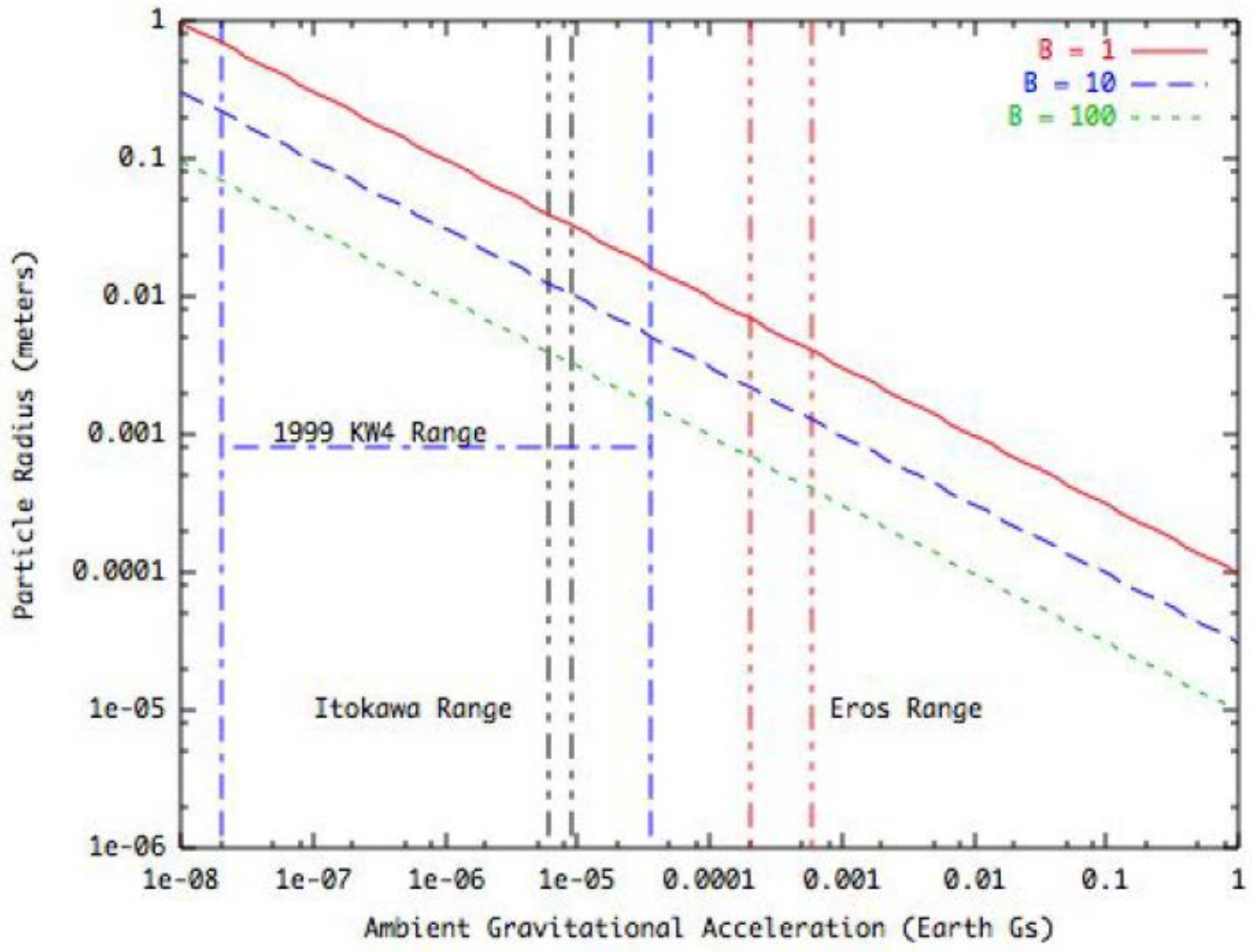

**Figure 16: Bond numbers as a function of particle radius and gravitational acceleration (from Scheeres et al. 2010). Ryugu's ambient gravity is estimated of the order of 0.25 milli G.**

As inter-grain cohesional forces act on the grain friction level, they act similarly to overburden pressure regarding macroscopic friction and grain rolling. This may easily be shown regarding the normal $F_N$ and tangential forces $F_T$ on grain level:

$$F_N = F_r - F_C \qquad (17)$$

$$F_T = (F_N + F_C)\mu \qquad (18)$$

where $F_r$ is the repulsive force due to contact, $F_C$ the total cohesion force and μ the micro-scale friction coefficient. The effect of cohesion lowers the repulsive effect of contact forces by holding grains together, whereas this effect does not lower the effective normal force for friction as cohesion keeps the contact in a tense condition. Regarding this issue, the magnitude of frictional force may not only depend on the level of gravity and the bond number, but also on the loading history. Higher loads that might have occurred in the past will compact the material establishing a larger number of bonds at possibly larger bond areas and thus increase the total amount of contact tension. The higher contact tension and thus micro-scale normal forces, the higher the friction forces and hence the macro-scale shear strength.

Given the expected angle of friction as stated above and the expected grain shape, inter-grain friction can also be expected to be higher than for most terrestrial sands. Thus the inter-grain friction angle for a spherical surrogate particle is expected to be in the range of 40-50 deg.

### 4.3.6. Regolith Bearing Strength

Assuming that the shape of an interacting device with the regolith is a square or a circle, the bearing capacity or strength of the regolith is defined as $\sigma = 1.3 * c * N_c$, where 1.3 is the shape factor (corresponding to a square or a circle), $c$ is the regolith cohesion, and $N_c$ is the bearing capacity factor. Assuming that cohesion is of the order of 100 Pa (see Sec. 4.3.5.) and that $N_c$ is in the range 6-20 depending on the actual angle of friction (0-33 degrees), then the bearing strength may be of the order of a few kPa or lower.

### 4.3.7. Whole Rock Properties

We use as a reference the Master's thesis of S. Shigaki (2016, Kobe University) for regolith with mm size range particles (Table 7); for bigger rocks, strength scaling may apply, resulting in weaker strength with increasing size. Note that the tensile strength measure gives similar or slightly higher values.

**Table 7: Crushing strength of various materials from Shigaki (2016).**

| Materials | Crushing strength (MPa) |
|---|---|
| Chondrule from Allende (CV3) | 7.7 ± 5.9 |
| Chondrule from Saratov (L4) | 9.4 ± 6.0 |
| Glass bead (1 mm size) | 223 ± 61 |
| Dunite (a few mm) | 13.1 ± 2.6 |
| Basalt (a few mm) | 16.9 ± 2.6 |
| Sandstone (a few mm) * | 3.4 ± 1.1 |

* Sandstone can be considered as an upper limit for the whole rock properties of Ryugu.

The compressive (unconfined) strength of sandstone is of the order of 50-100 MPa, while tensile strength is of the order of a few MPa (Ahrens 1995). However, the rock porosity of Ryugu may be higher (~40%) than the rock porosity of sandstone (~20%), and therefore the compressive and tensile strengths may be a factor 3 lower than that of sandstone, similar to Weibern tuff (Poelchau et al. 2014). This means that we can assume that the compressive strength could be of the order of 15-30 MPa and the tensile strength about 1 MPa or less.

### 4.3.8. Presence of Ponds and Their Properties

For sample return missions, very flat areas among the wide variety of asteroid surfaces are the most sought after since they are the safest places to touch down. As revealed by asteroid surfaces imaged at the cm-scale, ponds are good candidates (Veverka et al. 2001; Robinson et al. 2001; Cheng et al. 2002; Miyamoto et al. 2007). Since it is likely ponds are present on Ryugu's surface, it is essential to know if there is a difference between the material present in the ponds, and the material in other areas. Do our chondrite samples help us solve this issue?

Some fine-grained cognate lithologies in chondrites (Figure 17 and Figure 18) have been inferred to be sedimentary features indicating that they sample asteroid ponds. A cognate clast in the Vigarano CV3 chondrite, first identified by Johnson et al. (1990) appears to be a sample from such a pond, from a C-type asteroid. The clast consists mainly of micron-sized grains of olivine, arranged into layers with varying composition and porosity. The densest (most Fe-rich) olivine grains are in sections of layers with the lowest bulk porosity, which appear bright in BSE images. The most remarkable aspect of the clast are crossbeds. Zolensky et al. (2002, 2017) proposed that this clast formed by the processes of electrostatic grain levitation and subsequent seismic shaking. They then described similar clasts in another Vigarano sample, and in a section of Allende – all CV3 chondrites, and most recently in the LL3 chondrite NWA 8330 (Zolensky et al. 2017). These clasts consist mainly of olivine, whose compositional range is essentially identical to the host meteorite, although there is a striking predominance of heavy, Fe-rich olivine in the bulk clast relative to the bulk host. However, because of the compositional zoning within the pond deposit layers, the top most surface is depleted in Fe-rich, heavier olivine. Based on these observations, samples collected from the surfaces of ponds on Ryugu will *significantly* differ from the bulk asteroid by being finer-grained, metal poor, and silicates will be Fe-poor. Note that Itokawa samples were also depleted from typical LL chondrites by being metal depleted. Grain sizes in Ryugu regolith probably will extend down to the micron size, as was the case for Itokawa.

This interpretation is however challenged by a recent survey of Fe-rich secondary phases in CV chondrites (Ganino and Libourel 2017), which states that the phases constituting such similar networks of veinlets formed in reduced conditions near the iron-magnetite redox buffer at low aSiO2 (log(aSiO2) < −1) and moderate temperature (210–610 °C). The various lithologies in CV3 chondrites, i.e., CV3Red, CV3OxA, CV3OxB, including those present in Figures 17 and 18, and their diverse porosity and permeability (MacPherson and Krot 2014) are in good agreement with CV3 lithologies being variably altered crustal pieces coming from an asteroid percolated heterogeneously via porous flow of hydrothermal fluids. These hot, reduced and iron-rich fluids resemble pervasive, Darcy flow type, supercritical hydrothermal fluids, and consistent with textural settings of secondary phases, as scattered patches in the CV chondrite matrices or, when the fluid is channelized, as subtle veinlet networks. If this interpretation is correct, the permeability/porosity/cohesiveness of the unprocessed Ryugu surface, assuming CV (or CO) chondrites are relevant proxy, will depend on the percolation of hydrothermal fluids in the Ryugu's parent body, and their efficiency to precipitate secondary phases, which will in turn control the formation and evolution of the regolith, its size distribution and the ease of its sampling (see section 3).

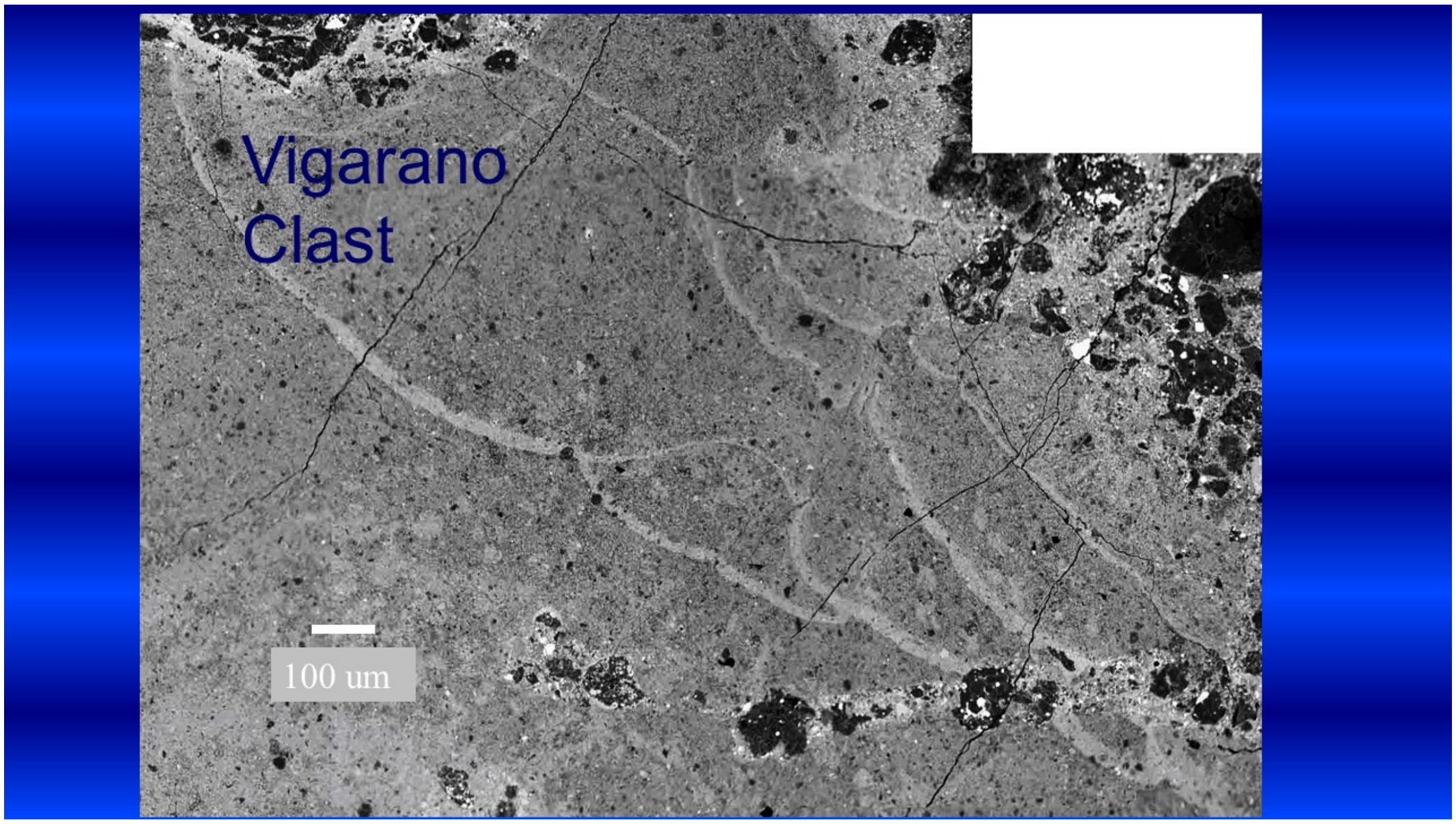

**Figure 17: Back-scattered electron image of a portion of a pond deposit clast from the Vigarano CV3 chondrite. The mineralogy of the clast is mainly olivine, with Fe-rich olivine being lighter in shade. Sedimentary beds are apparent, and light grey (Fe-olivine-rich) layers are at the bottoms of the beds. Crossbeds are apparent.**

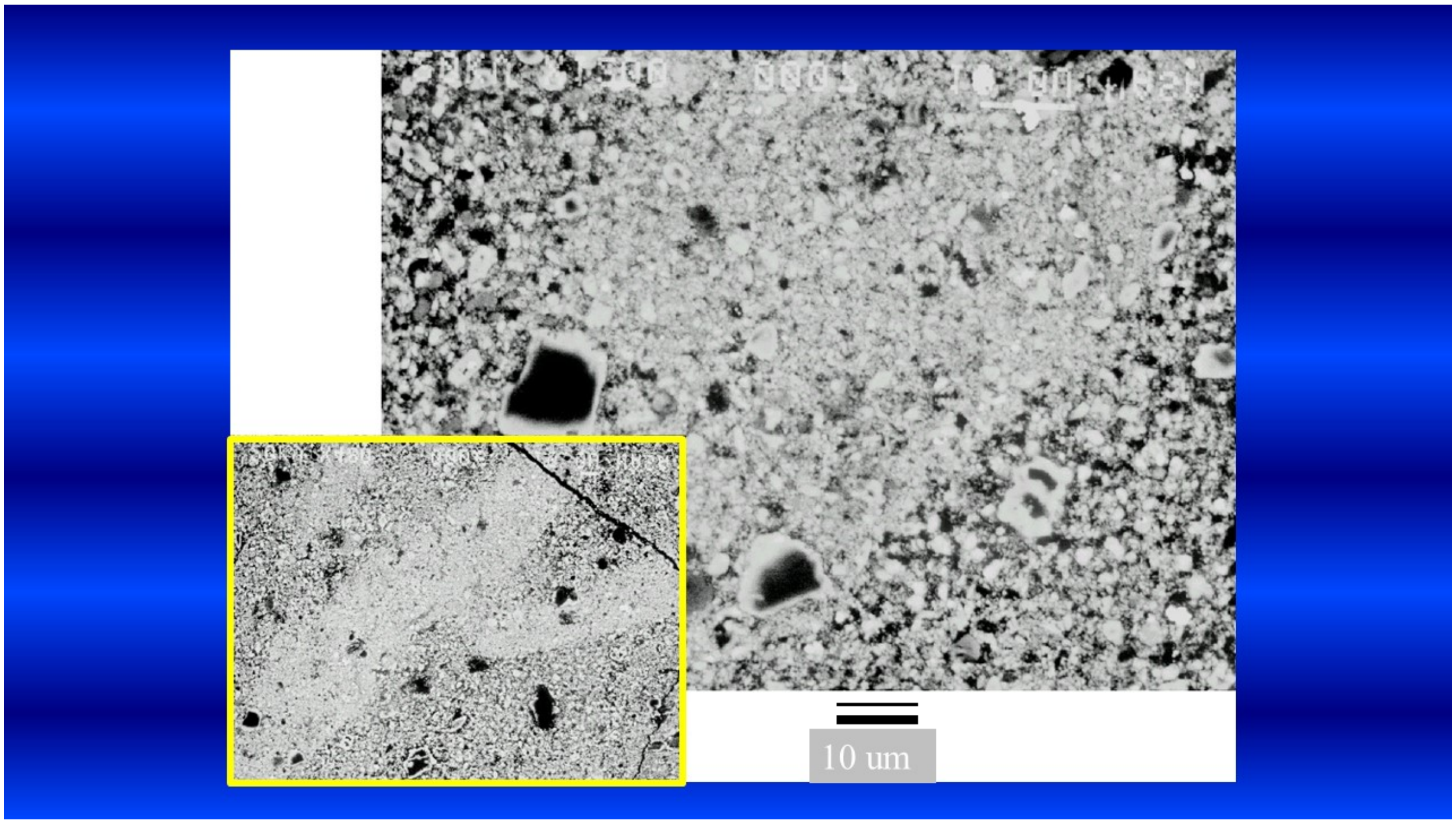

**Figure 18: Close-up on the clast in Figure 17, showing the bottom of one bed (upper left) with a sharp contact with the top of the succeeding bed (lower right). The mineralogy of the clast is mainly olivine, with Fe-rich olivine being lighter in shade. The highest concentration of light grey, Fe-rich olivine is at the bottom of the bed. Insert image at lower left is a lower mag view of the same area, with the high mag area at the lower left of the insert. Note that the grain size extends down to the micron scale.**

## 5. Summary and Conclusions

We have reviewed a wide range of information about the C-type, NEA (162173) Ryugu from the point of view of describing the regolith layer of Ryugu: the global properties, the thermophysical properties, and the mechanical properties. Each property is categorized into three types, i.e., the measured/observed, the derived, and the predicted properties, as summarized in Table 8. This table gives a reference model of Ryugu, especially on the surface regolith.

**Table 8: Properties of Ryugu, covered and discussed in the text.**

| | | | |
|---|---|---|---|
| Global properties | | | |
| Measured | Eccentricity | 0.190208 | |
| | Semi-major axis | 1.189555 | AU |
| | Inclination | 5.883556 | deg |
| | Period | 473.8878 | day |
| | Perihelion | 0.963292 | AU |
| | Aphelion | 1.415819 | AU |
| | Rotation period | 7.6326 | hour |
| | Pole direction in ecliptic coordinates (lambda, beta) | (310-340,-40 +/- 15) | deg |
| | Shape | almost spherical | |
| | Volume-equivalent diameter | 850-880 | m |
| | Phase function (Lommel-Seeliger model) | see Table 2 | |
| | Geometric and bond albedos | see Table 3 | |
| | Spectral type | C-type, see also Table 4 | |
| Derived | Composition (meteorite spectral counterpart) | heated CM, CM2, or CI | |
| Predicted | Satellite | no | |
| | Dust around Ryugu | no or little | |
| | Boulders: power law exponent of cumulative SFD | -3 | |
| | Crater density | more than Itokawa,<br>less than Toutatis | |
| | Surface roughness at meter scales | like Eros and Itokawa | |
| | Origin | originated in the inner Main Asteroid belt, between ~2.1-2.5 AU, and reached the $\nu_6$ by inward Yarkovsky drift | |
| Regolith thermophysical properties | | | |
| Measured | Thermal inertia | 150-300, typically 200 | J $m^{-2}$ $s^{-1/2}$ $K^{-1}$ |
| | Maximum surface temperature | 320-375 | K |
| | Roughness (the rms of surface slopes) | < 0.1 | |
| Derived | Regolith thermal conductivity | 0.020-0.108, most likely 0.042 | W $m^{-1}$ $K^{-1}$ |
| | Typical particle size in diameter | 3-30 most likely 6-10 (2.2-5 by Gundlach & Blum 2013 model) | mm |
| Predicted | Regolith heat capacity at 300 K | 758 | J $kg^{-1}$ $K^{-1}$ |
| | Regolith emissivity | 0.9 | |
| | Regolith thermal albedo | 0.019 | |
| | Regolith bulk density | 1100-1500 | kg $m^{-3}$ |
| Regolith mechanical properties | | | |
| Predicted | Regolith porosity | ~0.4-0.5 | |
| | Regolith cohesion | 100 | Pa |
| | Regolith angle of friction | 33 | deg |
| | Regolith tensile strength | < 1000 | Pa |
| | Inter-grain friction angle | 40-50 | deg |
| | Regolith bearing strength | < a few 1000 | Pa |
| | Compressive strength of rock | 15-30 | MPa |
| | Tensile strength of rock | < 1 | MPa |
| | Presence of ponds | possible, there compositions being expected to be significantly different from that of the bulk asteroid | |

As a whole, this paper provides a reference model of Ryugu's regolith which should be extremely useful both for science applications until Hayabusa2's arrival and for Hayabusa2 operations at Ryugu. It will be checked and refined thanks to Hayabusa2 data and further related scientific studies will be promoted toward understanding Ryugu, other small bodies, and Solar System evolution.

## 6. Acknowledgements

This paper is a product of IRSG in Hayabusa2 project, JAXA.

P.M. and M. A. B. acknowledge financial support from the French space agency CNES. M.E.Z. is supported by the NASA Emerging Worlds and Hayabusa2 Participating Scientist Programs, and the SERVII Center for Lunar Science and Exploration.

## 7. References

Abe, M.; et al., 2008, Ground-based Observational Campaign for Asteroid 162173 1999 JU3, 39th Lunar Planet. Sci., 1594

T. J. Ahrens (eds.), Rock Physics and phase relations: a handbook of physical constants, 1995.

T. Arai, T. Nakamura, S. Tanaka, H. Demura, Y. Ogawa, N. Sakatani, Y. Horikawa, H. Senshu, T. Fukuhara, T. Okada, Thermal imaging performance of TIR onboard the Hayabusa2 spacecraft. Space Sci. Rev. 208, 239-254 (2017).

M. Arakawa, K. Wada, T. Saiki, T. Kadono, Y. Takagi, K. Shirai, C. Okamoto, H. Yano, M. Hayakawa, S. Nakazawa, N. Hirata, M. Kobayashi, P. Michel, M. Jutzi, H. Imamura, K. Ogawa, N. Sakatani, Y. Iijima, R. Honda, K. Ishibashi, H. Hayakawa, H. Sawada, Scientific objectives of Small Carry-on Impactor (SCI) and Deployable Camera 3 Digital (DCAM3-D): observation of an ejecta curtain and a crater formed on the surface of Ryugu by an artificial high-velocity impact. Space Sci. Rev. 208, 187-212 (2017).

Barnouin-Jha, O.S., Cheng, A.F., Mukai, T., Abe, S., Hirata, N., Nakamura, R., Gaskell, R.W., Saito, J., Clark, B.E., 2008. Small-scale topography of 25143 Itokawa from the Hayabusa laser altimeter. Icarus 198, 108–124. doi:10.1016/j.icarus.2008.05.026

Benner, L.A.M., Ostro, S.J., Magri, C., Nolan, M.C., Howell, E.S., Giorgini, J.D., Jurgens, R.F., Margot, J.-L., Taylor, P.A., Busch, M.W., Shepard, M.K., 2008. Near-Earth asteroid surface roughness depends on compositional class. Icarus 198, 294–304. doi:10.1016/j.icarus.2008.06.010

Benner, L.A.M., Busch, M.W., Giorgini, J.D., Taylor, P.A., Margot, J.-L., 2015. Radar observations of near-Earth and main-belt asteroids. In: Asteroids IV.

P. Beck, A. Maturilli, A. Garenne, P. Vernazza, J. Helbert, E. Quirico, B. Schmitt, What is controlling the reflectance spectra (0.35- 150 µm) of hydrated (and dehydrated) carbonaceous chondrites? Submitted to Icarus (2018).

J.-P. Bibring, V. Hamm, Y. Langevin, C. Pilorget, A. Arondel, M. Bouzit, M. Chaigneau, B. Crane, A. Darié, C. Evesque, J. Hansotte, V. Gardien, L. Gonnod, J.-C. Leclech, L. Meslier, T. Redon, C. Tamiatto, S. Tosti, N. Thoores, The MicrOmega investigation onboard Hayabusa2. Space Sci. Rev. 208, 401-412 (2017).

Binzel, R.P., Harris, A.W., Bus, S.J., Burbine, T.H., 2001. Spectral properties of near-Earth objects: Palomar and IRTF results for 48 objects including spacecraft targets (9969) Braille and (10302)1989 ML. Icarus 151, 139–149.

R. P. Binzel, F. E. DeMeo, B. J. Burt, E. A. Cloutis, B. Rozitis, T. H. Burbine, H. Campins, B. E. Clark, J. P. Emery, C. W. Hergenrother, E. S. Howell, D. S. Lauretta, M. C. Nolan, M. Mansfield, V. Pietrasz, D. Polishook, D. J. Scheeres, Spectral slope variations for OSIRIS-REx target Asteroid (101955) Bennu: Possible evidence for a fine-grained regolith equatorial ridge, Icarus 256, 22-29 (2015).

Bottke W. F., Morbidelli A., Jedicke R., Petit J-M., Levison H., Michel P., Metcalfe T., "Debiased Orbital and Absolute Magnitude Distribution of the Near-Earth Objects" 2002, Icarus, 156, 399.

Bottke W. F., and 9 co-authors, "In search of the source of asteroid (101955) Bennu: Applications of the stochastic YORP model" 2015, Icarus, 247, 191.

Bowell, E., 1989, Application of photometric models to asteroids, Asteroids II; Proceedings of the Conference, Tucson, AZ, Mar. 8-11, 1988 (A90-27001 10-91), University of Arizona Press, 1989, p. 524-556.

Britt, D. T. and G. J. Consolmagno (2001). "Modeling the Structure of High Porosity Asteroids." Icarus 152(1): 134-139.

Bus S.J., Binzel R.P., 2002. Phase II of the small main-belt asteroid spectroscopic survey: A feature-based taxonomy. Icarus 158, 146–177.

Campins H., de Leon J., Morbidelli A., Licandro J., Gayon-Markt J., Delbo M., and Michel P., "The Origin of Asteroid 162173 (1999 JU3)" 2013 AJ, 146, 26.

Capaccioni, F., Cerroni, P., Coradini, M., Di Martino, M., Farinella, P., 1986. Asteroidal catastrophic collisions simulated by hypervelocity impact experiments. Icarus (ISSN 0019-1035) 66, 487–514. doi:10.1016/0019-1035(86)90087-4

Chapman, C.R., Merline, W.J., Thomas, P.C., Joseph, J., Cheng, A.F., Izenberg, N., 2002. Impact History of Eros: Craters and Boulders. Icarus 155, 104–118. doi:10.1006/icar.2001.6744

A. F. Cheng, Near Earth asteroid rendezvous: Mission summary. In: Bottke, W.F., Cellino, A., Paolicchi, P., Binzel, R.P. (Eds.), Asteroids III. Univ. Arizona Press, Tucson, pp. 351–366 (2002).

Cheng, A.F., Barnouin-Jha, O., Prockter, L., Zuber, M.T., Neumann, G., Smith, D.E., Garvin, J., Robinson, M., Veverka, J., Thomas, P.C., 2002. Small-Scale Topography of 433 Eros from Laser Altimetry and Imaging. Icarus 155, 51–74. doi:10.1006/icar.2001.6750.

B. E. Clark, J. Veverka, P. Helfenstein, P. C. Thomas, J. F. Bell, A. Harch, M. S. Robinson, S. L. Murchie, L. A. McFadden, C. R. Chapman, NEAR Photometry of Asteroid 253 Mathilde, Icarus 140, 53-65 (1999).

Cloutis E. A., Hudon P., Hiroi T., Gaffey M. J., 2012, Spectral reflectance properties of carbonaceous chondrites 4: Aqueously altered and thermally metamorphosed meteorites, Icarus, 220, 586-617.

Consolmagno, G.J.,.Schaefer, M.W., Schaefer, B.E., Britt, D.T., Macke, R.J.,Nolan, M.C., Howell, E.S.,The measurement of meteorite heat capacity at low temperatures using liquid nitrogen vaporization, Planetary and Space Science, 87, 146–156, 2013.

Davidsson, B.J.R., Rickman, H., 2014. Surface roughness and three-dimensional heat conduction in thermophysical models. Icarus 243, 58–77.

Davidsson B.J.R., H. Rickman, J.L. Bandfield, O. Groussin, P.J. Gutiérrez, M. Wilska, M.T. Capria, J.P. Emery, J. Helbert, L. Jorda, A. Maturilli, T.G. Mueller, Interpretation of thermal emission. I. The effect of roughness for spatially resolved atmosphereless bodies, Icarus, 252, 1-21, 2015

DeMeo F.E., Binzel R.P., Slivan S.M., Bus S.J., 2009. An extension of the Bus asteroid taxonomy into the near-infrared. Icarus 202, 160–180.

de León, J and 15 co-authors, “Visible spectroscopy of the Polana-Eulalia family complex: Spectral homogeneity” 2016, Icarus, 266, 57.

M. Delbo, G. Libourel, J. Wilkerson, N. Murdoch, P. Michel, K. T. Ramesh, C. Ganino, C. Verati, S. Marchi, Thermal fatigue as the origin of regolith on small asteroids, Nature 508, 233-236 (2014).

Demura, H., Kobayashi, S., Nemoto, E., Matsumoto, N., Furuya, M., Yukishita, A., Muranaka, N., Moritz, H., Shirakawa, K., Maruya, M., Ohyama, H., Uo, M., Kubota, T., Hashimoto, T., Kawaguchi, J., Fujiwara, A., Saito, J., Sasaki, S., Miyamoto, H., Hirata, N., 2006. Pole and Global Shape of 25143 Itokawa. Science 312, 1347–1349. doi:10.1126/science.1126574

Durda, D.D., Campo Bagatin, A., Alemañ, R.A., Flynn, G.J., STRAIT, M.M., Clayton, A.N., Patmore, E.B., 2015. The shapes of fragments from catastrophic disruption events: Effects of target shape and impact speed. Planetary and Space Science 107, 77–83. doi:10.1016/j.pss.2014.10.006

Durda, D.D., Scheeres, D.J., Roark, S.E., Sanchez, D.P. " Asteroid Regolith Mechanical Properties: Laboratory Experiments With Cohesive Powders", American Astronomical Society, DPS meeting #44, id.105.04 (2012).

Ernst, C.M., Rodgers, D.J., Barnouin, O.S., Murchie, S.L., Chabot, N.L., 2015. Evaluating Small Body Landing Hazards Due to Blocks. 46th Lunar and Planetary Science Conference 46, 2095.

Fountain, J.A., and E.A. West, Thermal conductivity of particulate basalt as a function of density in simulated lunar and martian environment, Journal of Geophysical Research, 75, 4063--4069, 1970.

Fujii, N., Osako, M., Thermal Diffusivity of Lunar Rocks under Atmospheric and Vacuum Conditions, Earth and Planetary Science Letters, 18, 65-71, 1973

Fujiwara, A., Kamimoto, G., Tsukamoto, A., 1978. Expected shape distribution of asteroids obtained from laboratory impact experiments. Nature 272, 602. doi:10.1038/272602a0

Fujiwara, A., Kawaguchi, J., Yeomans, D.K., Abe, M., Mukai, T., Okada, T., Saito, J., Yano, H., Yoshikawa, M., Scheeres, D.J., Barnouin-Jha, O., Cheng, A.F., Demura, H., Gaskell, R.W., Hirata, N., Ikeda, H., Kominato, T., Miyamoto, H., Nakamura, A.M., Nakamura, R., Sasaki, S., Uesugi, K., 2006. The Rubble-Pile Asteroid Itokawa as Observed by Hayabusa. Science 312, 1330–1334. doi:10.1126/science.1125841

Ganino C., Libourel G., (2017) Reduced and unstratified crust in CV chondrite parent body. *Nature communications* 8: 261, Published online 2017 Aug 15. doi: 10.1038/s41467-017-00293-1

Giblin, I., Martelli, G., Farinella, P., Paolicchi, P., Di Martino, M., Smith, P.N., 1998. The Properties of Fragments from Catastrophic Disruption Events. Icarus 134, 77–112. doi:10.1006/icar.1998.5937

M. Grott, J. Knollenberg, B. Borgs, F. Hänschke, E. Kessler, J. Helbert, A. Maturilli, N. Müller, The MASCOT radiometer MARA for the Hayabusa 2 mission. Space Sci. Rev. 208, 413-431 (2017).

Grott, M., J. Knollenberg, and C. Krause, Apollo lunar heat flow experiment revisited: A critical reassessment of the in situ thermal conductivity determination, Journal of Geophysical Research, 115, E11005, 2010.

Gundlach, B. and J. Blum (2012). " Outgassing of icy bodies in the Solar System - II: Heat transport in dry, porous surface dust layers." Icarus 219, 618-629.

Gundlach, B. and J. Blum (2013). "A new method to determine the grain size of planetary regolith." Icarus 223(1): 479-492.

Halajian J. D. and J. Reichman (1969). Correlation of mechanical and thermal properties of the lunar surface, Icarus. 10, 179-196.

M. Hamm, H. Senshu, M. Grott, Latitudinal Dependence of Asteroid Regolith Formation by Thermal Fatigue, Icarus, (2018) submitted.

Hapke, B. (2012). Theory of Reflectance and Emittance Spectroscopy.  Cambridge University Press.

Hapke, B. and H. Sato (2016). "The porosity of the upper lunar regolith." Icarus 273: 75-83.

Harris, A. W. and L. Drube (2016). "Thermal tomography of asteroid surface structure." The Astrophysical Journal 832(2): 127.

Heiken, G., D. Vaniman, and B. French (Eds.) (1991), Lunar Sourcebook, Cambridge Univ. Press, Cambridge, U. K.

Hemingway, B.S., Robie, R.A., Wilson, W.H., Specific heats of lunar soils, basalt, and breccias from the Apollo 14, 15, and 16 landing sites, between 90 and 350 K, Proceedings of the fourth Lunar Science Conference, Vol. 3., pp 2481-2487  1973

D. Herčík, H.-U. Auster, J. Blum, K.-H. Fornaçon, M. Fujimoto, K. Gebauer, C. Güttler, O. Hillenmaier, A. Hördt, E. Liebert, A. Matsuoka, R. Nomura, I. Richter, B. Stoll, B.P. Weiss, K.-H. Glassmeier, The MASCOT magnetometer. Space Sci. Rev. 208, 433-449 (2017).

Hirata, N., Barnouin-Jha, O.S., Honda, C., Nakamura, R., Miyamoto, H., Sasaki, S., Demura, H., Nakamura, A.M., Michikami, T., Gaskell, R.W., Saito, J., 2009. A survey of possible impact structures on 25143 Itokawa. Icarus 200, 486–502. doi:10.1016/j.icarus.2008.10.027

Hiroi T., Zolensky M. E., Pieters C. M., Lipschutz M. E., 1996. Thermal metamorphism of the C, G, B, and F asteroids seen from the 0.7 µm, 3 µm, and UV absorption strengths in comparison with carbonaceous chondrites, Meteoritics & Planetary Science, 31(3), 321-327

Hiroi T., Moroz L.V., Shingareva T.V., Basilevsky A.V., Pieters C.M., 2003. Effects of microsecond pulse laser irradiation on vis–NIR reflectance spectrum of carbonaceous chondrite stimulant: Implications for martian moons and primitive asteroids. 34th Lunar Planet. Sci., 1324.

T.-M. Ho, V. Baturkin, C. Grimm, J.T. Grundmann, C. Hobbie, E. Ksenik, C. Lange, K. Sasaki, M. Schlotterer, M. Talapina, N. Temtanasombat, E. Wejmo, L. Witte, M. Wrasmann, G. Wübbles, J. Rößler, C. Ziach, R. Findlay, J. Biele, C. Krause, S. Ulamec, M. Lange, O. Mierheim, R. Lichtenheldt, M. Maier, J. Reill, H.-J. Sedlmayr, P. Bousquet, A. Bellion, O. Bompis, C. Cenac-Morthe, M. Deleuze, S. Fredon, E. Jurado, E. Canalias, R. Jaumann, J.-P. Bibring, K.H. Glassmeier, D. Herčík, M. Grott, L. Celotti, F. Cordero, J. Hendrikse, T. Okada, MASCOT—The Mobile Asteroid Surface Scout onboard the Hayabusa2 mission. Space Sci. Rev. 208, 339-374 (2017).

Huang, J., Ji, J., Ye, P., Wang, X., Yan, J., Meng, L., Wang, S., Li, C., Li, Y., Qiao, D., Zhao, W., Zhao, Y., Zhang, T., Liu, P., Jiang, Y., Rao, W., Li, S., Huang, C., Ip, W.-H., Hu, S., Zhu, M., Yu, L., Zou, Y., Tang, X., Li, J., Zhao, H., Huang, H., Jiang, X., Bai, J., 2013. The Ginger-shaped Asteroid 4179 Toutatis: New Observations from a Successful Flyby of Chang'e-2. Nature Scientific Reports 3, 3411. doi:10.1038/srep03411

Hütter, E. S., N. I. Koemle, G. Kargl, and E. Kaufmann (2008), Determination of the effective thermal conductivity of granular materials under varying pressure conditions, J. Geophys. Res., 113, E12004, doi:10.1029/2008JE003085.

K. Ishibashi, K. Shirai, K. Ogawa, K. Wada, R. Honda, M. Arakawa, N. Sakatani, Y. Ikeda, Performance of Hayabusa2 DCAM3-D camera for short-range imaging of SCI and ejecta curtain generated from the artificial impact crater formed on asteroid 162137 Ryugu (1999 $JU_3$ ). Space Sci. Rev. 208, 213-238 (2017).

Ishiguro, M., Kuroda, D., Hasegawa, S. et al., 2014, Optical Properties of (162173) 1999 JU3: In Preparation for the JAXA Hayabusa 2 Sample Return Mission, ApJ 792, 74.

T. Iwata, K. Kitazato, M. Abe, M. Ohtake, T. Arai, T. Arai, N. Hirata, T. Hiroi, C. Honda, N. Imae, M. Komatsu, T. Matsunaga, M. Matsuoka, S. Matsuura, T. Nakamura, A. Nakato, Y. Nakauchi, T. Osawa, H. Senshu, Y. Takagi, K. Tsumura, N. Takato, S. Watanabe, M.-A. Barucci, E. Palomba, M. Ozaki, NIRS3: the near infrared spectrometer on Hayabusa2. Space Sci. Rev. 208, 317-337 (2017).

R. Jaumann, N. Schmitz, A. Koncz, H. Michaelis, S.E. Schroeder, S. Mottola, F. Trauthan, H. Hoffmann, T. Roatsch, D. Jobs, J. Kachlicki, B. Pforte, R. Terzer, M. Tschentscher, S. Weisse, U. Müller, L. Perez-Prieto, B. Broll, A. Kruselburger, T.-M. Ho, J. Biele, S. Ulamec, C. Krause, M. Grott, J.-P. Bibring, S. Watanabe, S.

Sugita, T. Okada, M. Yoshikawa, H. Yabuta, The camera of the MASCOT asteroid lander onboard Hayabusa2. Space Sci. Rev. 208, 375-400 (2017).

Jiang, Y., Ji, J., Huang, J., Marchi, S., Li, Y., Ip, W.-H., 2015. Asteroid 4179 Toutatis: boulders distribution as closely flew by Chang'e-2. IAU General Assembly 22.

Johnson K. L., K. Kendall, and A. D. Roberts (1971). Surface energy and the contact of elastic solids, Proc. R. Soc. Lond. A. Meth. Phys. Sci. 324, 301–313.

C. A. Johnson, M. Prinz, M. K. Weisberg, R. N. Clayton, T. K. Mayeda, Dark inclusions in Allende, Leoville, and Vigarano - Evidence for nebular oxidation of CV3 constituents, Geochimica et Cosmochimica Acta 54, 819-830 (1990).

S. Kameda, H. Suzuki, Y. Cho, S. Koga, M. Yamada, T. Nakamura, T. Hiroi, H. Sawada, R. Honda, T. Morota, C. Honda, A. Takei, T. Takamatsu, Y. Okumura, M. Sato, T. Yasuda, K. Shibasaki, S. Ikezawa, S. Sugita, Detectability of hydrous minerals using ONC-T camera onboard the Hayabusa2 spacecraft. Adv. Space Res. 56, 1519–1524 (2015).

S. Kameda, H. Suzuki, T. Takamatsu, Y. Cho, T. Yasuda, M. Yamada, H. Sawada, R. Honda, T. Morota, C. Honda, M. Sato, Y. Okumura, K. Shibasaki, S. Ikezawa, S. Sugita, Preflight calibration test results for optical navigation camera telescope (ONC-T) onboard the Hayabusa2 spacecraft. Space Sci. Rev. 208, 17-31 (2017).

M.-J. Kim, Y.-J. Choi, H.-K. Moon, M. Ishiguro, S. Mottola, M. Kaplan, D. Kuroda, D. S. Warjurkar, J. Takahashi, Y.-I. Byun, Optical observations of NEA 162173 (1999 JU3) during the 2011-2012 apparition, Astronomy & Astrophysics, 550, L11, 4 pp (2013).

M.-J. Kim, Y.-J. Choi, H.-K. Moon, Y. N. Krugly, S. Greenstreet, T. Lister, S. Kaynar, V. V. Rumyantsev, I. E. Molotov, Z. Donchev, Optical observations of NEA 162173 Ryugu (1999 JU3) during the 2016 apparition, American Astronomical Society, DPS meeting #48, 326.12 (2016).

King, A.J., Maturilli, A., Schofield1, P.F., Helbert, J., and Russell, S.S, Thermal Alteration of CI and CM Chondrites: Links to Primitive C-Type Asteroid Surfaces. Hayabusa Symposium, 2016.

Kiuchi, M. and A. M. Nakamura (2014). "Relationship between regolith particle size and porosity on small bodies." Icarus 239(0): 291-293.

Kleinhans, M.G., Markies, H., de Vet, S.J., et al. (2011) " Static and dynamic angles of repose in loose granular materials under reduced gravity", Journal of Geophysical Research: Planets, 116

Küppers, M., Moissl, R., Vincent, J.-B., Besse, S., Hviid, S.F., Carry, B., Grieger, B., Sierks, H., Keller, H.U., Marchi, S., Team, O., 2012. Boulders on Lutetia. Planetary and Space Science 66, 71–78. doi:10.1016/j.pss.2011.11.004

Langseth, M.G., S.J. Keihm, and K. Peters, Revised lunar heat-flow values, Proceedings of Lunar Science Conference, 7, 3143–3171, 1976.

D. Lazzaro, M. A. Barucci, D. Perna, F. L. Jasmim, M. Yoshikawa, J. M. F. Carvano, Rotational spectra of (162173) 1999 JU3, the target of the Hayabusa2 mission, Astron. Astrophys. 549, L2, 4 pp (2013).

D.S. Lauretta, S.S. Balram-Knutson, E. Beshore, W.V. Boynton, C. Drouet d'Aubigny, D.N. DellaGiustina, H.L. Enos, D.R. Gholish, C.W. Hergenrother, E.S. Howell, C.A. Johnson, E.T. Morton, M.C. Nolan, B. Rizk, H.L. Roper, A.E. Bartels, B.J. Bos, J.P. Dworkin, D.E. Highsmith, M.C. Moreau, D.A. Lorenz, L.F. Lim, R. Mink, J.A. Nuth, D.C. Reuter, A.A. Simon, E.B. Bierhaus, B.H. Bryan, R. Ballouz, O.S. Barnouin, R.P. Binzel, W.F. Bottke, V.E. Hamilton, K.J. Walsh, S.R. Chesley, P.R. Christensen, B.E. Clark, H.C. Connolly, M.K. Crombie, M.G. Daly, J.P. Emery, T.J. McCoy, J.W. McMahon, D.J. Scheeres, S. Messenger, K. Nakamura-Messenger, K. Righter, S.A. Sandford, OSIRIS-REx: sample return from asteroid (101955) Bennu. Space Sci. Rev. 212(1–2), 925–984 (2017).

L. Le Corre, J. A. Sanchez, V. Reddy, D. Takir, E. A. Cloutis, A. Thirouin, K. J. Becker, J.-Y. Li, S. Sugita, E. Tatsumi, Ground-based characterization of Hayabusa2 mission target asteroid 162173 Ryugu: constraining mineralogical composition in preparation for spacecraft operations, Monthly Notices of the Royal Astronomical Society 475, 614–623 (2018).

Ledlow, M.J., Burns, J.O., Gisler, G.R., Zhao, J.-H., Zeilik, M., Baker, D.N., 1992, Subsurface Emission from Mercury: VLA Radio Observations at 2 and 6 Centimeters, *ApJ*, 384:640-655.

Lee, P., Veverka, J., Thomas, P.C., Helfenstein, P., Belton, M.J.S., Chapman, C.R., Greeley, R., Pappalardo, R.T., Sullivan, R., Head, J.W.I., 1996. Ejecta Blocks on 243 Ida and on Other Asteroids. Icarus 120, 87–105. doi:10.1006/icar.1996.0039

Lee, S.W., Thomas, P., Veverka, J., 1986. Phobos, Deimos, and the Moon: Size and distribution of crater ejecta blocks. Icarus 68, 77–86. doi:10.1016/0019-1035(86)90075-8.

MacPherson, G. J., Krot, A. N., (2014) The formation of Ca-, Fe-rich silicates in reduced and oxidized CV chondrites: the roles of impact-modified porosity and permeability, and heterogeneous distribution of water ices. Meteorit. Planet. Sci. 49, 1250–1270.

J.-L. Margot, P. Pravec, P. Taylor, B. Carry, S. Jacobson, Asteroid Systems: Binaries, Triples, and Pairs, In: Bottke, W.F., Cellino, A., Paolicchi, P., Binzel, R.P. (Eds.), Asteroids III. Univ. Arizona Press, Tucson, 355–374 (2015).

Marsset, M., Carry, B., Dumas, C., Hanuš, J., Viikinkoski, M., Vernazza, P., Müller, T.G., Delbo, M., Jehin, E., Gillon, M., Grice, J., Yang, B., Fusco, T., Berthier, J., Sonnett, S., Kugel, F., Caron, J., Behrend, R., 3D shape of asteroid (6) Hebe from VLT/SPHERE imaging: Implications for the origin of ordinary H chondrites, A&A, 604 (2017) A64, doi:10.1051/0004-6361/201731021

Matsumura, S., Richardson, D.C., Michel, P., Schwartz, S.R., Ballouz, R.-L. 2014. The Brazil nut effect and its application to asteroids. Month. Not. Roy. Astron. Soc. 443, 3368-3380.

Maurel, C., Ballouz, R.L., Richardson, D.C, Michel, P., Schwartz, S.R. Numerical simulations of oscillation-driven regolith motion: Brazil-nut effect. Month. Not. Roy. Astron. Soc. 464, 2866-2881, (2017).

Mazrouei, S., Daly, M.G., Barnouin, O.S., Ernst, C.M., DeSouza, I., 2014. Block distributions on Itokawa. Icarus 229, 181–189. doi:10.1016/j.icarus.2013.11.010

Michel, P., Richardson, D.C. 2013. Collision and gravitational reaccumulation: Possible formation mechanism of the asteroid Itokawa. Astron. Astrophys. 554, L1-L4.

Michikami, T., Hagermann, A., Kadokawa, T., Yoshida, A., Shimada, A., Hasegawa, S., Tsuchiyama, A., 2016. Fragment shapes in impact experiments ranging from cratering to catastrophic disruption. Icarus 264, 316–330. doi:10.1016/j.icarus.2015.09.038

Michikami, T., Nakamura, A.M., Hirata, N., 2010. The shape distribution of boulders on Asteroid 25143 Itokawa: Comparison with fragments from impact experiments. Icarus 207, 277–284. doi:10.1016/j.icarus.2009.10.008

Michikami, T., Nakamura, A. M., Hirata, N., Gaskell, R. W., Nakamura, R., Honda, T., Honda, C., Hiraoka, K., Saito, J., Demura, H., Ishiguro, M., Miyamoto, H., 2008. Size-frequency statistics of boulders on global surface of asteroid 25143 Itokawa. Earth, Planets and Space, 60, 13-20.

H. Miyamoto, H. Yano, D. J. Scheeres, S. Abe, O. Barnouin-Jha, A. F. Cheng, H. Demura, R. W. Gaskell, N. Hirata, M. Ishiguro, T. Michikami, A. M. Nakamura, R. Nakamura, J. Saito, S. Sasaki, Regolith Migration and Sorting on Asteroid Itokawa, Science 316, 1011-1014 (2007).

T. Mizuno, T. Kase, T. Shiina, M. Mita, N. Namiki, H. Senshu, R. Yamada, H. Noda, H. Kunimori, N. Hirata, F. Terui, Y. Mimasu, Development of the laser altimeter (LIDAR) for Hayabusa2. Space Sci. Rev. 208, 33-47 (2017).

Moskovitz N.A., Ane S., Pan K., Osip D.J., Pefkou D., Melita M.D., Elias M., Kitazato K., Bus S.J., DeMeo F.E., Binzel R.P., Abell P.A., 2013. Rotational characterization of Hayabusa II target Asteroid (162173) 1999 JU3, Icarus 224, 24-31.

Müller, T. G., J. Ďurech, M. Ishiguro, M. Müller, T. Krühler, H. Yang, M.-J. Kim, L. O'Rourke, F. Usui, C. Kiss, B. Altieri, B. Carry, Y.-J. Choi, M. Delbo, J. P. Emery, J. Greiner, S. Hasegawa, J. L. Hora, F. Knust, D. Kuroda, D. Osip, A. Rau, A. Rivkin, P. Schady, J. Thomas-Osip, D. Trilling, S. Urakawa, E. Vilenius, P. Weissman and P. Zeidler (2017). "Hayabusa-2 mission target asteroid 162173 Ryugu (1999 JU3): Searching for the object's spin-axis orientation⋆." A&A 599: A103.

Murdoch N., Sanchez, P., Schwartz, S.R., Miyamoto, H. (2015) "Asteroid Surface Geophysics", Asteroids IV, pp.767-792

Nakamura, T., T. Noguchi, M. Tanaka, M. E. Zolensky, M. Kimura, A. Tsuchiyama, A. Nakato, T. Ogami, H. Ishida, M. Uesugi, T. Yada, K. Shirai, A. Fujimura, R. Okazaki, S. A. Sandford, Y. Ishibashi, M. Abe, T. Okada, M. Ueno, T. Mukai, M. Yoshikawa and J. Kawaguchi (2011). "Itokawa Dust Particles: A Direct Link Between S-Type Asteroids and Ordinary Chondrites." Science 333(6046): 1113-1116.

K. Nagao, R. Okazaki, T. Nakamura, Y. N. Miura, T. Osawa, K. Bajo, S. Matsuda, M. Ebihara, T. R. Ireland, F. Kitajima, H. Naraoka, T. Noguchi, A. Tsuchiyama, H. Yurimoto, M. E. Zolensky, M. Uesugi, K. Shirai, M. Abe, T. Yada, Y. Ishibashi, A. Fujimura, T. Mukai, M. Ueno, T. Okada, M. Yoshikawa, J. Kawaguchi, Irradiation History of Itokawa Regolith Material Deduced from Noble Gases in the Hayabusa Samples, Science 333, 1128-1131 (2011).

N. Namiki, T. Mizuno, N. Hirata, H. Noda, H. Senshu, R. Yamada, H. Ikeda, S. Abe, K. Matsumoto, S. Oshigami, M. Shizugami, F. Yoshida, N. Hirata, H. Miyamoto, S. Sasaki, H. Araki, S. Tazawa, Y. Ishihara, M. Kobayashi, K. Wada, H. Demura, J. Kimura, M. Hayakawa, N. Kobayashi, Scientific use of LIDAR data of Hayabusa-2 mission, Proceeding of an International CJMT-1, Workshop on Asteroidal Science, 74-96 (2014).

Noviello, J.L., Ernst, C.M., Barnouin, O.S., Daly, M., 2014. Block Distribution on Itokawa: Implications for Asteroid Surface Evolution. 44th Lunar and Planetary Science Conference 45, 1587.

K. Ogawa, K. Shirai, H. Sawada, M. Arakawa, R. Honda, K. Wada, K. Ishibashi, Y. Iijima, N. Sakatani, S. Nakazawa, H. Hayakawa, System configuration and operation plan of Hayabusa2 DCAM3-D camera system for scientific observation during SCI impact experiment. Space Sci. Rev. 208, 125-142 (2017).

T. Okada, T. Fukuhara, S. Tanaka, M. Taguchi, T. Imamura, T. Arai, H. Senshu, Y. Ogawa, H. Demura, K. Kitazato, R. Nakamura, T. Kouyama, T. Sekiguchi, S. Hasegawa, T. Matsunaga, T. Wada, J. Takita, N. Sakatani, Y. Horikawa, K. Endo, J. Helbert, T.G. Müller, A. Hagermann, Thermal infrared imaging experiments of C-Type asteroid 162173 Ryugu on Hayabusa2. Space Sci. Rev. 208, 255-286 (2017).

T. Omura, A. M. Nakamura, Experimental study on compression property of regolith analogues, Planet. Space. Sci. 149, 14-22, (2017).

Pajola, M., Oklay, N., La Forgia, F., Giacomini, L., Massironi, M., Bertini, I., El-Maarry, M.R., Marzari, F., Preusker, F., Scholten, F., Höfner, S., Lee, J.-C., Vincent, J.-B., Groussin, O., Naletto, G., Lazzarin, M., Barbieri, C., Sierks, H., Lamy, P., Rodrigo, R., Koschny, D., Rickman, H., Keller, H.U., Agarwal, J., A'Hearn, M.F., Barucci, M.A., Bertaux, J.-L., Cremonese, G., Da Deppo, V., Davidsson, B., De Cecco, M., Debei, S., Ferri, F., Fornasier, S., Fulle, M., Güttler, C., Gutiérrez, P.J., Hviid, S.F., Ip, W.-H., Jorda, L., Knollenberg, J., Kramm, J.R., Küppers, M., Kürt, E., Lara, L.M., Lin, Z.-Y., Lopez Moreno, J.J., Magrin, S., Michalik, H., Mottola, S., Thomas, N., Tubiana, C., 2016. Aswan site on comet 67P/Churyumov-Gerasimenko:

Morphology, boulder evolution, and spectrophotometry. A&A 592, A69. doi:10.1051/0004-6361/201527865

Pajola, M., Vincent, J.-B., Güttler, C., Lee, J.-C., Bertini, I., Massironi, M., Simioni, E., Marzari, F., Giacomini, L., Lucchetti, A., Barbieri, C., Cremonese, G., Naletto, G., Pommerol, A., El-Maarry, M.R., Besse, S., Küppers, M., La Forgia, F., Lazzarin, M., Thomas, N., Auger, A.-T., Sierks, H., Lamy, P., Rodrigo, R., Koschny, D., Rickman, H., Keller, H.U., Agarwal, J., A'Hearn, M.F., Barucci, M.A., Bertaux, J.-L., Da Deppo, V., Davidsson, B., De Cecco, M., Debei, S., Ferri, F., Fornasier, S., Fulle, M., Groussin, O., Gutiérrez, P.J., Hviid, S.F., Ip, W.-H., Jorda, L., Knollenberg, J., Kramm, J.R., Kürt, E., Lara, L.M., Lin, Z.-Y., Lopez Moreno, J.J., Magrin, S., Marchi, S., Michalik, H., Moissl, R., Mottola, S., Oklay, N., Preusker, F., Scholten, F., Tubiana, C., 2015. Size-frequency distribution of boulders ≥7 m on comet 67P/Churyumov-Gerasimenko. A&A 583, A37. doi:10.1051/0004-6361/201525975

Perna D., Barucci M.A., Ishiguro M., Alvarex-Candal A., Kuroda D., Yoshiwaka M., Kim M., Fornasier S., Hasegawa S., Roh D., Müller T.G., Kim Y., 2017. Spectral and rotational properties of near-Earth asteroid (162173) Ryugu, target of the Hayabusa2 sample return mission, A&A 599, L1.

Pinilla-Alonso N., Lorenzi V., Campins H., de Leon J., Licandro J., 2013. Near-infrared spectroscopy of 1999 JU3, the target of the Hayabusa 2 mission, A&A 552, A79.

Pinilla-Alonso, N. and 10 co-authors, "Portrait of the Polana-Eulalia family complex: Surface homogeneity revealed from near-infrared spectroscopy" 2016 Icarus, 274, 231.

Piqueux S. and P. R. Christensen (2009). A model of thermal conductivity for planetary soils: 1. Theory for unconsolidated soils, Journal of Geophysical Research, 114, E09005.

Piqueux, S., and P. R. Christensen (2011), Temperaturedependent thermal inertia of homogeneous Martian regolith, J. Geophys. Res., 116, E07004, doi:10.1029/2011JE003805.

Poelchau, M. H., Kenkmann, T., Hoerth, T., Schafer, F., Rudolf, M., Thoma, K., 2014. Impact cratering experiments into quartzite, sandstone and tuff: The effects of projectile size and target properties on spallation. Icarus 242, 211-224.

Rivkin A.S., Howell E.S., Vilas F., Lebofksy L.A., 2002. Hydrated minerals on asteroids: The astronomical record. In: Bottke, W.F., Cellino, A., Paolicchi, P., Binzel, R.P. (Eds.), Asteroids III. Univ. Arizona Press, Tucson, pp. 235–253.

Robie, R.A., Hemingway, B.S., Wilson, W.H., Specific Heats of Lunar Surface Materials from 90 to 350K, Proceedings of the Apollo 11 Lunar Science Conference, Vol. 3, pp 2361-2367, 1970

M. S. Robinson, P. C. Thomas, J. Veverka, S. Murchie, B. Carcich, The nature of ponded deposits on Eros, Nature 413, 396-400 (2001).

Robinson, M.S., Thomas, P.C., Veverka, J., Murchie, S.L., Wilcox, B.B., 2002. The geology of 433 Eros. Meteoritics & Planetary Science 37, 1651–1684. doi:10.1111/j.1945-5100.2002.tb01157.x

Rodgers, D.J., Ernst, C.M., Barnouin, O.S., Murchie, S.L., Chabot, N.L., 2016. Methodology for Finding and Evaluating Safe Landing Sites on Small Bodies. Planetary and Space Science 1–37. doi:10.1016/j.pss.2016.10.010

Rousseau, E., A. Siria, G. Jourdan, S. Volz, F. Comin, J. Chevrier and J.-J. Greffet (2009). "Radiative heat transfer at the nanoscale." Nature Photonics 3(9): 514-517.

B. Rozitis, S. F. Green, Directional characteristics of thermal-infrared beaming from atmosphereless planetary surfaces - a new thermophysical model, Monthly Notices of the Royal Astronomical Society 415, 2042-2062 (2011).

Rozitis, B., Green, S.F., MacLennan, E., Emery, J.P., The Variations of Asteroid Thermal Inertia with Heliocentric Distance, Asteroids, Comets, Meteors, Montevideo, 2017.

T. Saiki, H. Imamura, M. Arakawa, K. Wada, Y. Takagi, M. Hayakawa, K. Shirai, H. Yano, C. Okamoto, The small carry-on impactor (SCI) and the Hayabusa2 impact experiment. Space Sci. Rev. 208, 165-186 (2017).

Saito, J., Miyamoto, H., Nakamura, R., Ishiguro, M., Michikami, T., Nakamura, A.M., Demura, H., Sasaki, S., Hirata, N., Honda, C., Yamamoto, A., Yokota, Y., Fuse, T., Yoshida, F., Tholen, D.J., Gaskell, R.W., Hashimoto, T., Kubota, T., Higuchi, Y., Nakamura, T., Smith, P., Hiraoka, K., Honda, T., Kobayashi, S., Furuya, M., Matsumoto, N., Nemoto, E., Yukishita, A., Kitazato, K., Dermawan, B., Sogame, A., Terazono, J., Shinohara, C., Akiyama, H., 2006. Detailed Images of Asteroid 25143 Itokawa from Hayabusa. Science 312, 1341–1344. doi:10.1126/science.1125722

Sakatani, N., Ogawa K., Iijima Y., Arakawa M., Tanaka S., Compressional stress effect on thermal conductivity of powdered materials: Measurements and their implication to lunar regolith, Icarus, Volume 267, 1-11, http://dx.doi.org/10.1016/j.icarus.2015.12.012, 2016.

Sakatani N., K. Ogawa, Y. Iijima, M. Arakawa, R. Honda, and S. Tanaka (2017). Thermal conductivity model for powdered materials under vacuum based on experimental studies, AIP Advances, 7, 015310.

Sakatani N., K. Ogawa, M. Arakawa, and S. Tanaka (2018). Thermal conductivity of lunar regolith simulant JSC-1A under vacuum, Icarus, 309, 13-24, https://doi.org/10.1016/j.icarus.2018.02.027.

H. Sawada, K. Ogawa, K. Shirai, S. Kimura, Y. Hiromori, Y. Mimasu (DCAM3 Development Team), Deployable Camera (DCAM3) system for observation of Hayabusa2 impact experiment. Space Sci. Rev. 208, 143-164 (2017).

Scheeres, D.J., Hartzell C.M., Sanchez, P. (2010). „Scaling forces to asteroid surfaces: The role of cohesion", Icarus 210 (2), 968-984

Schräpler, R., J. Blum, I. von Borstel and C. Güttler (2015). "The stratification of regolith on celestial objects." Icarus 257: 33-46.

Schreiner, S. S., J. A. Dominguez, L. Sibille and J. A. Hoffman (2016). "Thermophysical property models for lunar regolith." Advances in Space Research 57(5): 1209-1222.

Schröder, S. E., Mottola, S., Carsenty, U., Ciarniello, M., Juamann, R., Li, J.-Y., Longobardo, A., Plamer, E., Pierters, C., Preusker, F., Raymond, C. A., Russel, C. T. Resolved spectrophotometric properties of the Ceres surface from Dawn Framing Camera images. (2017) Icarus 288: 201-225.

H. Senshu, S. Oshigami, M. Kobayashi, R. Yamada, N. Namiki, H. Noda, Y. Ishihara, T. Mizuno, Dust detection mode of the Hayabusa2 LIDAR. Space Sci. Rev. 208, 65-79 (2017).

H. Sierks, C. Barbieri, P.L. Lamy, R. Rodrigo, D. Koschny, H. Rickman, H.U. Keller, J. Agarwal, M.F. A'Hearn, F. Angrilli, A.-T. Auger, M.A. Barucci, J.-L. Bertaux, I. Bertini, S. Besse, D. Bodewits, C. Capanna, G. Cremonese, V. Da Deppo, B. Davidsson, S. Debei, M. De Cecco, F. Ferri, S. Fornasier, M. Fulle, R. Gaskell, L. Giacomini, O. Groussin, P. Gutierrez-Marques, P.J. Gutiérrez, C. Güttler, N. Hoekzema, S.F. Hviid, W.-H. Ip, L. Jorda, J. Knollenberg, G. Kovacs, J.R. Kramm, E. Kührt, M. Küppers, F. La Forgia, L.M. Lara, M. Lazzarin, C. Leyrat, J.J. Lopez Moreno, S. Magrin, S. Marchi, F. Marzari, M. Massironi, H. Michalik, R. Moissl, S. Mottola, G. Naletto, N. Oklay, M. Pajola, M. Pertile, F. Preusker, L. Sabau, F. Scholten, C. Snodgrass, N. Thomas, C. Tubiana, J.-B. Vincent, K.-P. Wenzel, M. Zaccariotto, M. Pätzold, On the nucleus structure and activity of comet 67P/Churyumov-Gerasimenko. Science 347, 1044 (2015).

H. Sierks, P. Lamy, C. Barbieri, D. Koschny, H. Rickman, R. Rodrigo, M. F. A'Hearn, F. Angrilli, M. A. Barucci, J.-L. Bertaux, I. Bertini, S. Besse, B. Carry, G. Cremonese, V. Da Deppo, B. Davidsson, S. Debei, M. De Cecco, J. De Leon, F. Ferri, S. Fornasier, M. Fulle, S. F. Hviid, R. W. Gaskell, O. Groussin, P. Gutierrez, W. Ip, L. Jorda, M. Kaasalainen, H. U. Keller, J. Knollenberg, R. Kramm, E. Kührt, M. Küppers, L. Lara, M Lazzarin, C. Leyrat, J. J. Lopez Moreno, S. Magrin, S. Marchi, F. Marzari, M. Massironi, H. Michalik, R. Moissl, G. Naletto, F. Preusker, L. Sabau, W. Sabolo, F. Scholten, C. Snodgrass, N. Thomas, C. Tubiana, P. Vernazza, J.-B. Vincent, K.-P. Wenzel, T. Andert, M. Pätzold, B. P. Weiss, Images of Asteroid 21 Lutetia: A Remnant Planetesimal from the Early Solar System. Science 334, 487-490 (2011).

S. Shigaki, Compression process of chondrite parent bodies inferred from the strength of chondrules (in Japanese), Master's thesis, Kobe University (2016).

Singh, B. P. and M. Kaviany (1994). "Effect of solid conductivity on radiative heat transfer in packed beds." International Journal of Heat and Mass Transfer 37(16): 2579-2583.

Stebbins, J., Carmichael, I., Moret, L., 1984. Heat capacities and entropies of silicate liquids and glasses. Contrib. Mineral. Petrol. 86 (2), 131–148.

Sugita S. et al., 2013, Visible Spectroscopic Observations of Asteroid 162173 (1999 JU3) with the Gemini-S Telescope, 44th Lunar Planet. Sci., 2591

Susorney, H.C., Barnouin, O.S., 2016. The Global Surface Roughness of 433 Eros from the NEAR-Shoemaker Laser Altimeter (NLR). AAS/Division for Planetary Sciences Meeting Abstracts 48.

Suzuki M., K. Makino, M. Yamada, and K. Iinoya (1980). A study of coordination number in a random packed system of monosized sphere particles (in Japanese), Kagaku Kogaku Ronbunshu 6, 59–64.

Takir D., Clark B.E., d'Aubigny C.D., Hergenrother C.W., Li J.Y., Lauretta D.S., and Binzel R.P. (2015). Photometric models of disk-integrated observations of the OSIRIS-REx target Asteroid (101955) Bennu. *Icarus*, 252, 393-399.

J. Takita, H. Senshu, S. Tanaka, Feasibility and accuracy of thermophysical estimation of asteroid 162173 Ryugu (1999 JU3) from the Hayabusa2 thermal infrared imager. Space Sci. Rev. 208, 287-315 (2017).

Tholen D., Barucci M. A. 1989, in Asteroids II, edn. R. P. Binzel, T. Gehrels, & M. S. Matthews (Univ. of Arizona Press), 298

Thomas, P.C., Robinson, M.S., 2005. Seismic resurfacing by a single impact on the asteroid 433 Eros. Nature 436, 366–369. doi:10.1038/nature03855

Thomas, P.C., Veverka, J., Bell, J.F., Clark, B.E., Carcich, B., Joseph, J., Robinson, M., McFadden, L.A., Malin, M.C., Chapman, C.R., Merline, W., Murchie, S., 1999. Mathilde: Size, Shape, and Geology. Icarus 140, 17–27. doi:10.1006/icar.1999.6121

Thomas, P.C., Veverka, J., Robinson, M.S., Murchie, S., 2001. Shoemaker crater as the source of most ejecta blocks on the asteroid 433 Eros. Nature 413, 394–396.

Thomas, P.C., Veverka, J., Sullivan, R., Simonelli, D.P., Malin, M.C., Caplinger, M., Hartmann, W.K., James, P.B., 2000. Phobos: Regolith and ejecta blocks investigated with Mars Orbiter Camera images. Journal of Geophysical Research 105, 15091–15106. doi:10.1029/1999JE001204

Thuillet, F., Michel, P., Maurel, C., Ballouz, R.-L., Zhang, Y., Richardson, D.C., Biele, J. 2018. Numerical modeling of lander interaction with a low-gravity asteroid regolith surface : Application to MASCOT onboard Hayabusa2. Astron. Astrophys., submitted.

Tonui E., Zolensky M., Hiroi T., Nakamura T., Lipschutz M., Wang M.-S., Okudaira K. (2014) Petrographic, chemical and spectroscopic evidence for thermal metamorphism in carbonaceous chondrites I: CI and CM chondrites. Geochimica et Cosmochimica Acta 126, 284–306.

Tsuchiyama, A., M. Uesugi, K. Uesugi, T. Nakano, R. Noguchi, T. Matsumoto, J. Matsuno, T. Nagano, Y. Imai, A. Shimada, A. Takeuchi, Y. Suzuki, T. Nakamura, T. Noguchi, M. Abe, T. Yada and A. Fujimura (2014). "Three-dimensional microstructure of samples recovered from asteroid 25143 Itokawa: Comparison with LL5 and LL6 chondrite particles." Meteoritics & Planetary Science 49(2): 172-187.

Y. Tsuda, M. Yoshikawa, M. Abe, H. Minamino, S. Nakazawa, System design of the Hayabusa2—asteroid sample return mission to Ryugu. Acta Astronaut. 91, 356 (2013).

Turcotte, D.L., 1997. Fractals and chaos in geology and geophysics. Cambridge University Press, Cambridge, 398p.

J. Veverka, P. C. Thomas, M. Robinson, S. Murchie, C. Chapman, M. Bell, A. Harch, W. J. Merline, J. F. Bell, B. Bussey, B. Carcich, A. Cheng, B. Clark, D. Domingue, D. Dunham, R. Farquhar, M. J. Gaffey, E. Hawkins, N. Izenberg, J. Joseph, R. Kirk, H. Li, P. Lucey, M. Malin, L. McFadden, J. K. Miller, W. M. Owen, C. Peterson, L. Prockter, J. Warren, D. Wellnitz, B. G. Williams, D. K. Yeomans, Imaging of Small-Scale Features on 433 Eros from NEAR: Evidence for a Complex Regolith, Science 292, 484-488 (2001).

Vilas F., 2008. Spectral characteristics of Hayabusa 2 near-Earth asteroid targets 162173 1999 JU3 and 2001 QC34, The Astronomical Journal, 135: 1101-1105.

K. J. Walsh, S. A. Jacobson, Formation and evolution of binary asteroids, In: P. Michel, F. E. DeMeo, W. F. Bottke (Eds.), Asteroids IV, Univ. Arizona Press, Tucson, pp. 375-393 (2015).

S. Watanabe, Y. Tsuda, M. Yoshikawa, S. Tanaka, T. Saiki, S. Nakazawa, Hayabusa2 mission overview, 2017, Space Sci. Rev. 208, 3-16, DOI 10.1007/s11214-017-0377-1.

Wechsler, A.E., Glaser, P.E., Fountain, J.A., Thermal properties of granulated materials, Prog. Aero. Astron. 28, 215 (1972).

D. F. Winter, J. M. Saari, A Particulate Thermophysical Model of the Lunar Soil, Astrophys. J. 156, 1135-1151 (1969).

R. Yamada, H. Senshu, N. Namiki, T. Mizuno, S. Abe, F. Yoshida, H. Noda, N. Hirata, S. Oshigami, H. Araki, Y. Ishihara, K. Matsumoto, Albedo observation by Hayabusa2 LIDAR: instrument performance and error evaluation. Space Sci. Rev. 208, 49-64 (2017).

Yomogida, K., Matsui, T., Physical properties of ordinary chondrites, Journal of Geophysical Research, 88, 9513-9533, doi:10.1029/JB088iB11p09513, 1983.

M. Yoshikawa, J. Kawaguchi, A. Fujiwara, A. Tsuchiyama, Hayabusa Sample Return Mission, In: P. Michel, F. E. DeMeo, W. F. Bottke (Eds.)., Asteroids IV, Univ. Arizona Press, Tucson, pp. 397-418 (2015).

M. E. Zolensky, K. Nakamura, A. F. Cheng, M. J. Cintala, F. Hörz, R. V. Morris, D. Criswell, Meteoritic Evidence for the Mechanism of Pond Formation on Asteroid Eros, Lunar and Planetary Science Conference 33 (2002).

M. Zolensky, L. Le, Asteroid pond mineralogy: view from a cognate clast in LL3 northwest africa 8330, Abstracts of the 2017 Meteoritical Society Meeting, #6164 (2017).

R. Zou, M. Gan, A. Yu, Prediction of the porosity of multi-component mixtures of cohesive and non-cohesive particles, Chemical engineering science 66(20), 4711-4721 (2011).

Zou, X., Li, C., Liu, J., Wang, W., Li, H., Ping, J., 2014. The preliminary analysis of the 4179 Toutatis snapshots of the Chang'E-2 flyby. Icarus 229, 348–354. doi:10.1016/j.icarus.2013.11.002

Zuber, M.T., Smith, D.E., Cheng, A.F., Garvin, J.B., Aharonson, O., Cole, T.D., Dunn, P.J., Guo, Y., Lemoine, F.G., Neumann, G.A., Rowlands, D.D., Torrence, M.H., 2000. The Shape of 433 Eros from the NEAR-Shoemaker Laser Rangefinder. Science 289, 2097–2101. doi:10.1126/science.289.5487.2097